\documentclass[reqno,12pt]{article}
\usepackage{amsthm,amsmath,amssymb,graphicx,setspace,verbatim,natbib}
\usepackage[small]{caption}
\usepackage{subcaption}
\usepackage{wrapfig}
\usepackage{multirow}
\usepackage{rotating}
\usepackage{color}
\usepackage{changepage}

\newcommand{\bzero}{\mbox{\boldmath $0$}}

\newcommand{\bb}{\mbox{\boldmath $b$}}

\newcommand{\bv}{\mbox{\boldmath $v$}}

\newcommand{\bw}{\mbox{\boldmath $w$}}
\newcommand{\bx}{\mbox{\boldmath $x$}}

\newcommand{\tby}{\tilde{\mbox{\boldmath $y$}}}
\newcommand{\by}{\mbox{\boldmath $y$}}
\newcommand{\bz}{\mbox{\boldmath $z$}}

\newcommand{\bH}{\mbox{\boldmath $H$}}
\newcommand{\bI}{\mbox{\boldmath $I$}}

\newcommand{\bM}{\mbox{\boldmath $M$}}

\newcommand{\bQ}{\mbox{\boldmath $Q$}}

\newcommand{\bS}{\mbox{\boldmath $S$}}

\newcommand{\bV}{\mbox{\boldmath $V$}}

\newcommand{\bX}{\mbox{\boldmath $X$}}
\newcommand{\bY}{\mbox{\boldmath $Y$}}

\newcommand{\cG}{{\cal G}}

\newcommand{\cI}{{\cal I}}

\newcommand{\cM}{{\cal M}}
\newcommand{\cN}{{\cal N}}

\newcommand{\cW}{{\cal W}}

\newcommand{\ty}{\tilde{y}}
\newcommand{\tY}{\tilde{Y}}

\newcommand{\tE}{\tilde{E}}

\newcommand{\eps}{\varepsilon}

\newcommand{\tbeta}{\tilde{\beta}}

\newcommand{\tsigma}{\tilde{\sigma}}

\newcommand{\teps}{\tilde{\eps}}

\newcommand{\balpha}{\mbox{\boldmath $\alpha$}}
\newcommand{\bbeta}{\mbox{\boldmath $\beta$}}

\newcommand{\btheta}{\mbox{\boldmath $\theta$}}

\newcommand{\bphi}{\mbox{\boldmath $\phi$}}

\newcommand{\bgamma}{\mbox{\boldmath $\gamma$}}

\newcommand{\bmu}{\mbox{\boldmath $\mu$}}
\newcommand{\tbmu}{\tilde{\mbox{\boldmath $\mu$}}}
\newcommand{\bpi}{\mbox{\boldmath $\pi$}}

\newcommand{\bSigma}{\mbox{\boldmath $\Sigma$}}

\newcommand{\tbQ}{\tilde{\mbox{\boldmath $Q$}}}

\newcommand{\bPsi}{\mbox{\boldmath $\Psi$}}

\newcommand{\bOmega}{\mbox{\boldmath $\Omega$}}
\newcommand{\bomega}{\mbox{\boldmath $\omega$}}
\newcommand{\bGamma}{\mbox{\boldmath $\Gamma$}}

\newcommand{\ups}{\upsilon}

\newcommand{\E}{\mbox{E}}
\newcommand{\Var}{\mbox{Var}}

\newcommand{\miss}{\mbox{\scriptsize miss}}

\newcommand{\trigamma}{\digamma'}
\newcommand{\pr}{\mbox{\scriptsize p}}

\newcommand{\bdm}{\begin{displaymath}}
\newcommand{\edm}{\end{displaymath}}
\newcommand{\beq}{\begin{equation}}
\newcommand{\eeq}{\end{equation}}

\renewcommand{\th}{^{\mbox{\scriptsize th}}}

\long\def\symbolfootnote[#1]#2{\begingroup%
\def\thefootnote{\fnsymbol{footnote}}\footnote[#1]{#2}\endgroup}

\newlength{\offsetpage}
{\end{adjustwidth}}

\begin{document}

\pagenumbering{arabic}

\pagestyle{empty}
\vspace{.1in}
\begin{center}
{\singlespacing
\begin{Large}{\bf
A Hierarchical Bayesian Model for Stochastic Spatiotemporal SIR Modeling and Prediction of COVID-19 Cases and Hospitalizations\\}
\vspace{.3in}
\end{Large}

Curtis B. Storlie, Ricardo L. Rojas, Gabriel O. Demuth, Benjamin D. Pollock, 
Patrick W. Johnson, Patrick M. Wilson, Ethan P. Heinzen, Hongfang Liu, 
Rickey E. Carter, Sean C. Dowdy, Shannon M. Dunlay, Elizabeth B. Habermann,
Daryl J. Kor, Matthew R. Neville, Andrew H. Limper, Katherine H. Noe,
Mohamad Bydon, Pablo Moreno Franco, Priya Sampathkumar, Nilay D. Shah,
Henry H. Ting.\\[.25in]

Mayo Clinic\\[.35in]

\begin{abstract}

  Most COVID-19 predictive modeling efforts use statistical or mathematical models to predict national- and state-level COVID-19 cases or deaths in the future. These approaches assume parameters such as reproduction time, test positivity rate, hospitalization rate, and social intervention effectiveness (masking, distancing, and mobility) are constant. However, the one certainty with the COVID-19 pandemic is that these parameters change over time, as well as vary across counties and states.  In fact, the rate of spread over region, hospitalization rate, hospital length of stay and mortality rate, the proportion of the population that is susceptible, test positivity rate, and social behaviors can all change significantly over time. Thus, the quantification of uncertainty becomes critical in making meaningful and accurate forecasts of the future.  Bayesian approaches are a natural way to fully represent this uncertainty in mathematical models and have become particularly popular in physics and engineering models.  The explicit integration time varying parameters and uncertainty quantification into a hierarchical Bayesian forecast model differentiates the Mayo COVID-19 model from other forecasting models such as YYG, MIT, IHME, CDC, Google, and others. By accounting for all sources of uncertainty in both parameter estimation as well as future trends with a Bayesian approach, the Mayo COVID-19 model accurately forecasts future cases and hospitalizations, as well as the degree of uncertainty. Additional features of the Mayo COVID-19 model presented here are automated updates to the model parameters and running the subsequent forecast simulations every day as new data are available, and incorporating leading indicators such as internet search trends (e.g. Google trends) and social mobility/proximity data (e.g. Unacast).  This approach has been remarkably accurate and a linchpin in Mayo Clinic's response to managing the COVID-19 pandemic.  The model accurately predicted timing and extent of the summer and fall surges at Mayo Clinic sites, allowing hospital leadership to manage resources effectively to provide a successful pandemic response.  This model has also proven to be very useful to the state of Minnesota to help guide difficult policy decisions.  When many other models were predicting very large rises in hospital census in the state of MN as a whole during the fall surge, our model accurately predicted a significant, but much much more modest rise and fall.  Two week case forecasts for each state and county from this model have now been made publicly available on the web.

\vspace{.15in}

\vspace{.15in}
\noindent
{\em Running title}: Stochastic SIR Modeling for COVID-19 Cases and Hospitalizations

\vspace{.15in}
\noindent
{\em Corresponding Author}: Curtis Storlie, \verb1storlie.curt@mayo.edu1

\end{abstract}
}

\end{center}


\setcounter{page}{0}
\newpage
\pagestyle{plain}
\vspace{-.35in}
\section{Introduction}
\vspace{-.05in}

\begin{figure}[b!]
\vspace{-.0in}
\centering
\caption{Trends in rate of spread and hospital admission rates for 4 selected states.}
\vspace{-.15in}
\includegraphics[width=.49\textwidth]{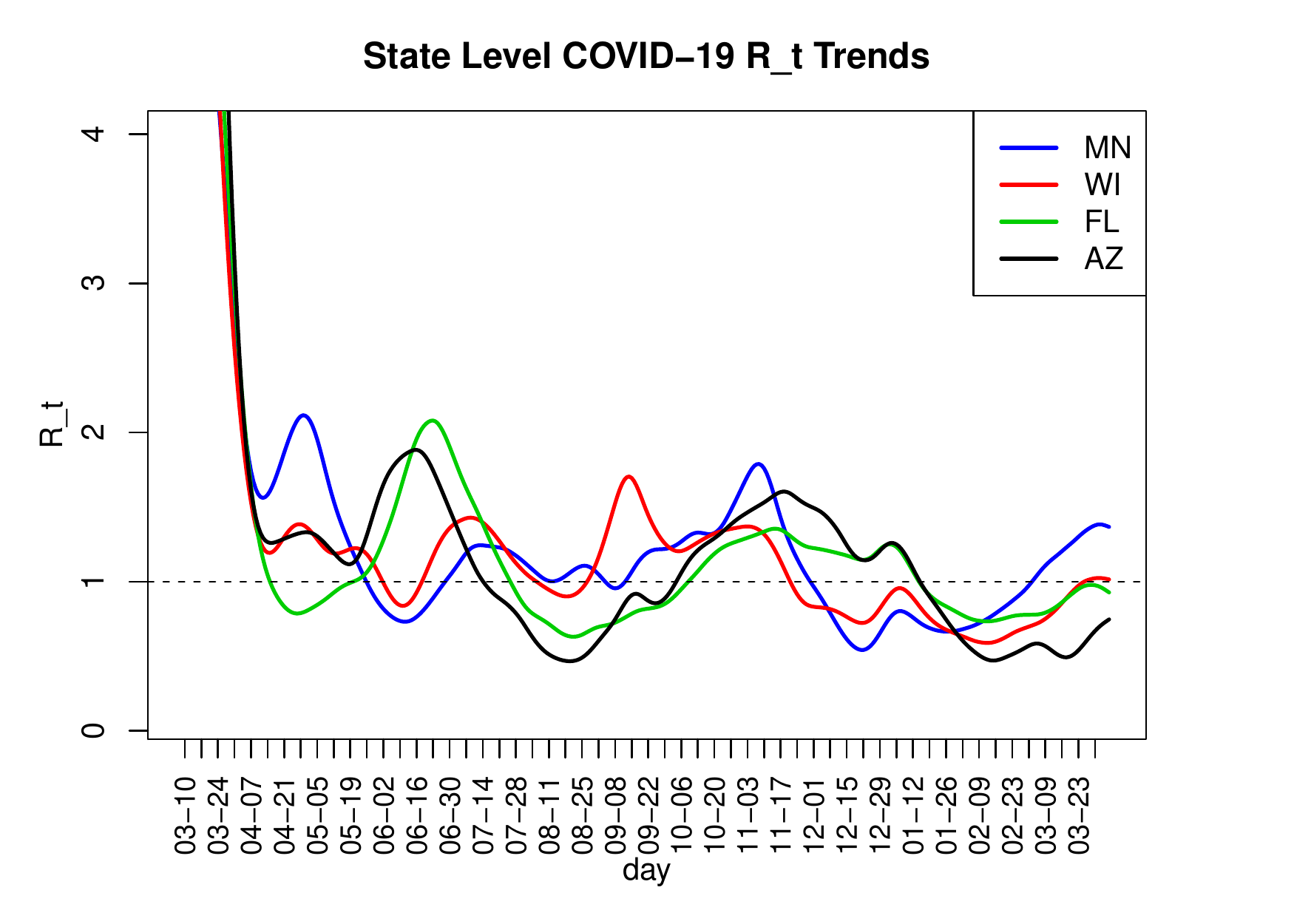}
\includegraphics[width=.49\textwidth]{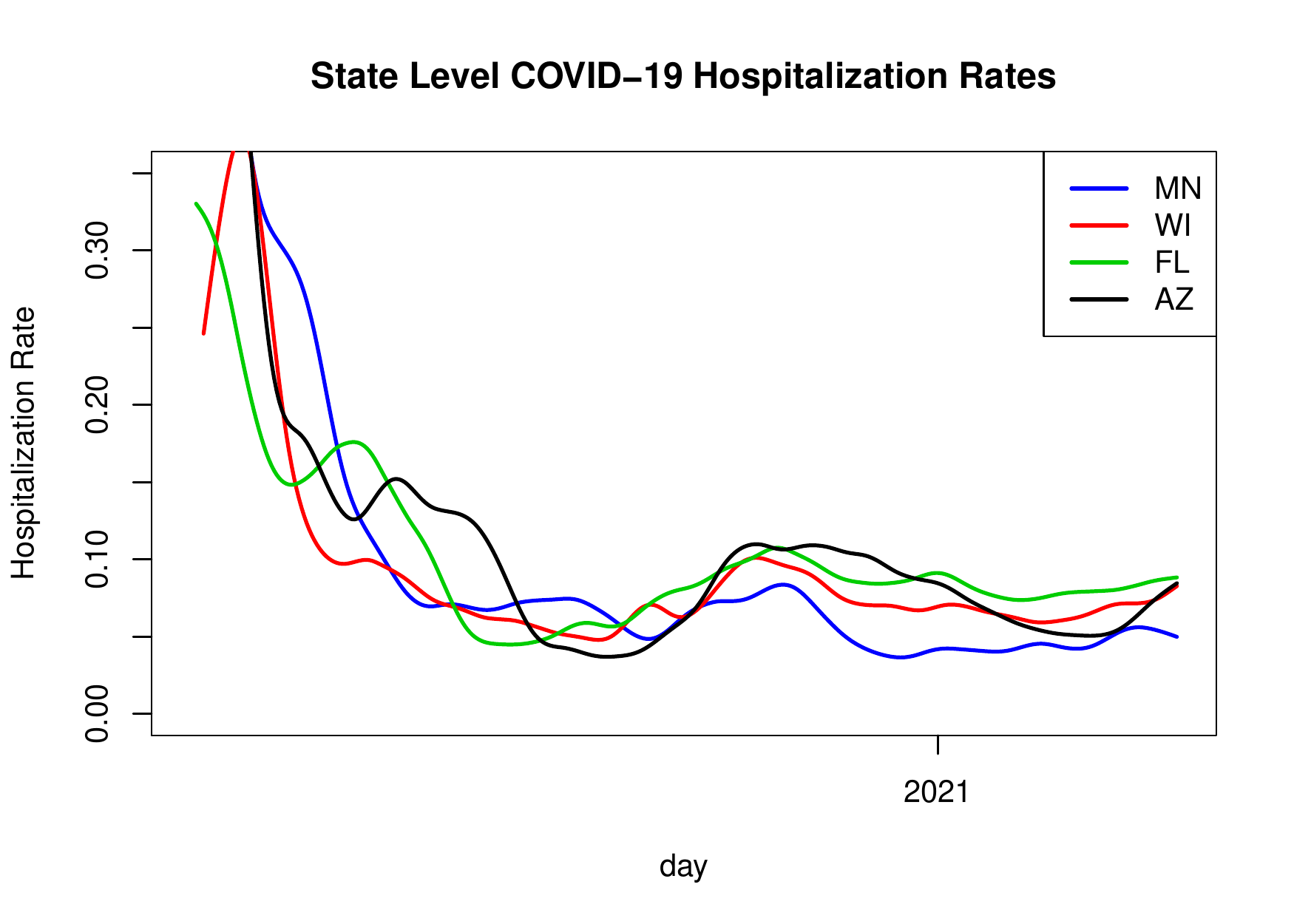}
\label{fig:trends}
\vspace{-.1in}
\end{figure}

The majority of the COVID-19 predictive modeling efforts use statistical or mathematical models to predict changes in national- and state-level cumulative reported COVID-19 cases or deaths in the future.  Those efforts leverage various types of data (e.g., COVID-19 data, demographic data, mobility data), methods (see below), and estimates of the impacts of interventions (e.g. social distancing, use of face coverings). However, there has been limited research on how best to incorporate time-varying information into risk predictions. The one certainty with the COVID-19 pandemic is that things have changed over time and will continue to do so, with respect to the rate of spread, the test positivity rate, the hospitalization rate, etc (see Figure~\ref{fig:trends}). These trends can be similar across regions, but not necessarily the same or with the same timing.

Social behavior, public policy, and disease dynamics like mutation play a large role in this variability in region and time.  It is thus naive to assume a traditional epidemiological model with constant parameters could characterize the true nature of the future.  A traditional Susceptible-Infected-Removed (SIR) model, for example, assumes the rate of transition from the susceptible to infected is proportional to the number of infected.  This can be transformed into an adequate model, provided the proportionality constant is allowed to change with region and time.  However, the rate of spread over region and time cannot possibly be known with any reasonable degree of certainty.  The same is true for other critical parameters such as hospitalization rate, the extent of the undercount of test positive cases compared to the true number of infections, and hospital length of stay related parameters. That is, uncertainty quantification (UQ) \citep{Helton06,Storlie09b,Reich09,Storlie12b} becomes critical in making any kind of meaningful forecast.  Bayesian approaches are a natural way to fully represent the uncertainty in mathematical models and have become very popular for this purpose in physics and engineering \citep{Kennedy01, Higdon04, Storlie2015calibration, Bhat17, Lai16, Storlie13onion}.

\begin{figure}[t!]
\vspace{-.13in}
\centering
\caption{Bayesian approach to model calibration and forecasting.}
\vspace{-.15in}
\includegraphics[width=.99\textwidth]{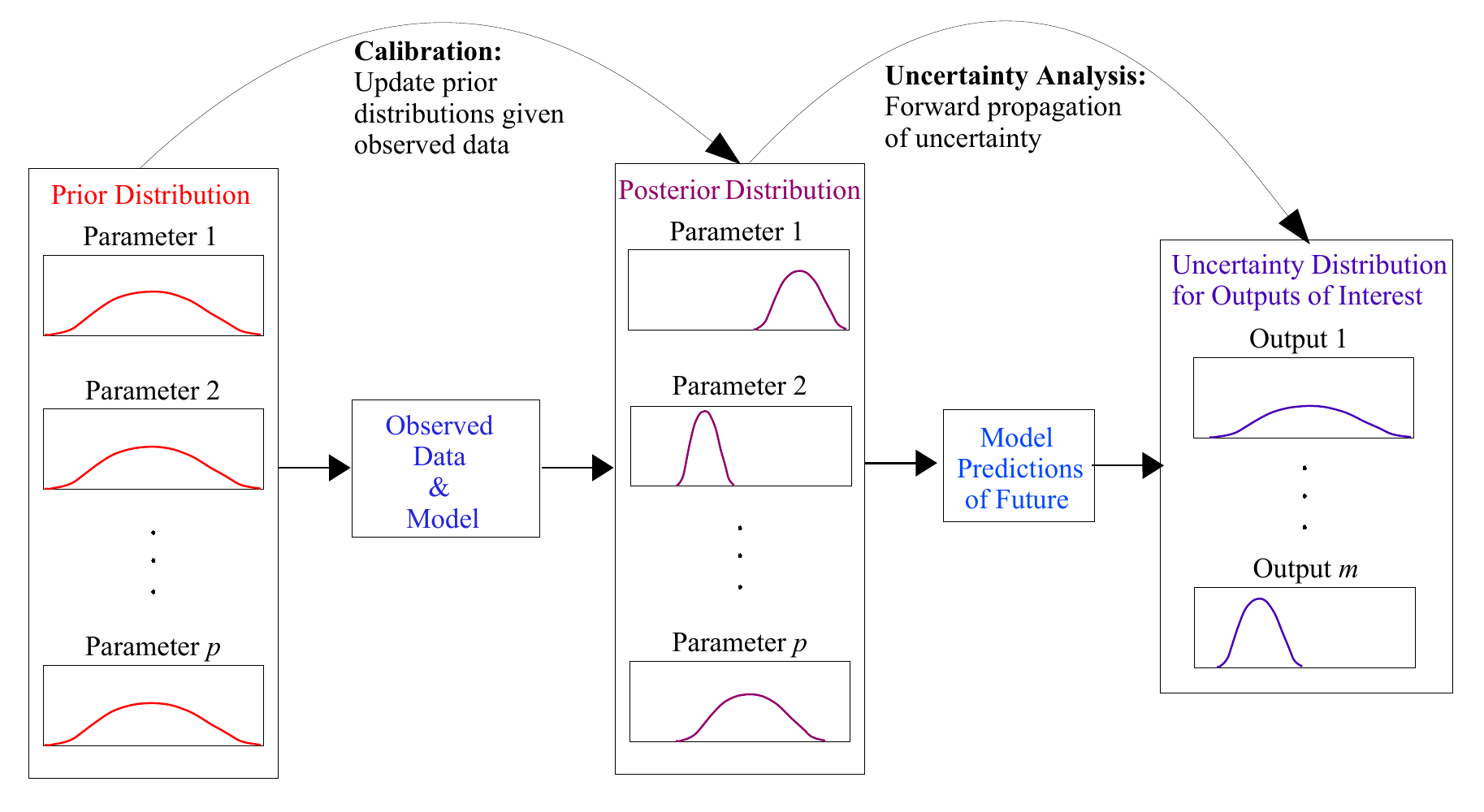}
\label{fig:bayes_UQ}
\vspace{-.1in}
\end{figure}

The Bayesian approach to model calibration, and forward propagation of uncertainty is described in Figure~\ref{fig:bayes_UQ}.  The underlying idea is that uncertainty is captured with a probability distribution.  There is a prior distribution for the model parameters or inputs, $\btheta$.  For a given parameter(s) this prior distribution can be informative, based on expert knowledge or evidence in literature, or it can be rather diffuse to represent a lack of knowledge for that parameter(s).  The prior distribution is then updated by conditioning on observed data, e.g., reported cases, number of tests, hospitalizations, etc.   The resulting conditional distribution is the posterior distribution which is generally approximated via Markov chain Monte Carlo (MCMC) sampling.  The process of obtaining this posterior distribution is called calibration.  Once obtained, predictions of the future can be made by running the model forward in time, while accounting for the uncertainty still present in $\btheta$ and represented by the posterior distribution.  Algorithmically, this can be conceptualized as taking a particular value of $\btheta$ from a sample of the posterior, using it as fixed to then make a model prediction, then repeating this process many times in a Monte Carlo fashion to obtain a sample of predictions.

There have been many models created and used for modeling COVID-19, including
MIT \citep{Kissler20}, Youyang Gu, a.k.a. YYG \citep{YYG}, IHME \citep{IHME20}, CDC \citep{ray2020ensemble}, LANL \citep{LANL}, and Google \citep{Arik2020}, among others.
The explicit integration of uncertainty quantification into a hierarchical Bayesian forecast model differentiates the Mayo COVID-19 model from most other COVID-19 forecasting models.  
\cite{LANL_sir} and \cite{lin2021} also apply the Bayesian UQ framework above to compartmental disease modeling and indeed this is the most similar existing work to the approach we take here.  The primary differences are in our explicit treatment of the undercount of infections (in a similar manner to YYG), the spatio-temporal trends at the county and day level, and the subsequent prediction of hospital census at the individual hospital level.

[Gabriel Paragraphs]

A great number of epidemic models have been used to attempt to predict the COVID-19 pandemic. \cite{zhou2020semiparametric} suggests dividing them into three basic groups; curve fitting models, compartment models, and agent based models. Following this typology, we compare our work to a number of other COVID-19 models. 

Curve fitting seems to be a relatively less popular method for forecasting. In two closely related works \cite{covid2020forecasting} and \cite{IHME20} use Gaussian error to fit a curve for COVID-19 hospitalizations. As part of their ensemble predictor, \cite{adiga2021all} utilize both autoregressive and Kalman Filter based curve fitting models. Both these types performed well relative to the other models in the predictor, with one variation of the autoregressive model including spatial effects. \cite{tamang2020forecasting} used an artificial neural net to do curve-fitting based forecasting of confirmed cases. \cite{YYG} takes a very high level approach to COVID modeling, based mostly around heuristics for converting between positive test rates and true infection counts. Thus, similar to our work, \cite{YYG} obtains estimates and projections of the true infected count, however, \cite{YYG} does not provide any form of uncertainty estimation, and in any case ceased providing estimates in March 2021 with the halt of the COVID Tracking Project \citep{covidtracking}.

Compartment models have a long history in disease modeling, first appearing in seminal work on SIR by \cite{kermack1991contributions} in 1927; see \cite{brauer2008lecture, bjornstad2020modeling} for a modern description of differential equation based modeling and \cite{allen1994some} for their discrete-time analogues. \cite{Kissler20} uses covariate-adjusted infection rates for longterm forecasting and exploration of possible future trajectories of COVID-19 out to 2024. Similar to our model, \cite{ihme2020modeling} allows the infection rate to vary with time, similar to the time varying infection rate in our model, in the previous citation infection rate is estimated via mixed effects, while other parameters that govern the dynamics of COVID-19 such as the mixing rate of infected individuals and the rates at which individuals transition from asymptomatic and from infected to recovered are sampled from distributions based on expert opinion. Final model uncertainty is quantified by repeatedly resampling these parameters and refitting the overall model \citep[Supplemental Material]{ihme2020modeling}. This method is essentially a partial implementation of a fully Bayesian model paradigm, since these unknown parameters are sampled from what amounts to their priors, rather than being informed by observed infection data within model fitting. Similar to our model, other authors have added new compartments to their models to count different outcomes or groups of interest. For instance a group at Google adds states for hospitalization, ICU admission, ventilator usage and death, a set of states very similar to that in our model \citep{Arik2020}. The Google model also adds states to account for undetected infections and accounts for uncertainty through quantile loss estimation, rather than a method that allows for uncertainty propagation. Another model fairly similar to this work is \cite{wang2020spatiotemporal}, which predicts reported infections at a state and county level in the United States which models geographic similarity through a bivariate spline surface that is allowed to vary over the spatial domain of the United States. However this model does not account for undetected infections. Because the model is fit under a frequentist paradigm, uncertainty is provided through bootstrap estimation. \cite{flaxman2020estimating} uses a Bayesian compartment model to estimate COVID-19 mortality in Europe, and allows for a time-varying infection rate, although this model does not include any explicit spatial correlation or structure. \cite{lin2021} uses a very complex compartment model based roughly on a SEIR structure, which differentiates not only between stages of infection, but between individuals practicing social distancing or not, as well as quarantined individuals. This model is fit using Bayesian methodology, and so produces uncertainty estimates fundamentally similar to our own. However, \cite{lin2021} does not account for spatial trends at all, and uses a static infection rate, which nevertheless varies at a population level based on the number of individuals in each compartment. 

\cite{ferguson2020report} utilizes agent-based modeling for COVID-19, based on earlier work for forecasting influenza epidemics \citep{halloran2008modeling, ferguson2006strategies}. These works assign socialization patterns to simulated individuals in a high-resolution geographic area, and allows for differing levels of contact at home, work, school, and the wider community. Because individual contacts are explicitly modeled, this approach allows for explicit counterfactuals such as the effect of canceling all large in-person gathers.  is based on earlier work modeling potential influenza outbreaks  This allows a detailed simulation of the effects of different governmental interventions on social contacts and the spread of the disease. Since individuals within the simulation are still in states such as susceptible and infected, this may be seen as an extension of SIR/SEIR or other state space models that allows for extreme heterogenaity within states, rather than assuming well mixed states. 

Finally, several groups have ensembled multiple different classes of models into a singe prediction. Beginning in April 2020, the Centers for Disease Control and Prevention (CDC) collected a number of predictive models developed by different teams using differing data and methodologies \citep{ray2020ensemble}. The only requirements for the included models is to produce quantile predictions on a set schedule; quantiles are then averaged together to produce the final ensemble prediction. More recently, \cite{adiga2021all} expands on the CDC ensemble predictor by fitting an autoregressive time series model, a Long Short Term Memory (LTSM) network, a SEIR model and a Kalman Filter based method, then ensembling the predictions using Bayesian model averaging. Although this utilizes a smaller number of models than \cite{ray2020ensemble}, Bayesian model averaging will put greater weight on models that correspond better to the data, unlike the CDC ensemble's naive equal-weight ensemble.  


In addition, the case counts driving the estimation of infection rate are a lagged version of the current state, i.e., positive cases today are really due to an increase in spread 7 days or longer ago
.

Thus, the short-term forecasts can also be improved using leading indicators such as search trends \citep{Kurian20,Carneiro2009} and mobility data \cite{Unacast,Yilmazkuday2020}.  Incorporating all of these data (case counts, hospital admissions, length of stay data, google search trends, mobility trends, etc.) some from county level, some from state level, some from specific hospitals, requires a very complex and intricate model, also begging for a Bayesian approach to tie it all together and represent all uncertainty in a principled manner.

In this paper we develop and demonstrate the use of a hierarchical Bayesian model for both short and long-term forecasting, and for state-level and local hospital census and staff absence projections.
The proposed model is a stochastic multi-state SIR model \citep{Xu2016b} for each U.S. county updated daily with case counts. The model state structure is provided in Figure 3.  The infection rate driving the SIR is a space-time, log-Gaussian process over each US county and day.  The proportion of people in a given county that are test-positive (test+) susceptible is also a critical parameter to the model for longer term forecasts.  This essentially captures how many people are truly infected as opposed to how many have had a positive test.  Thus, it is important to accurately represent this parameter and characterize its uncertainty. Since this parameter is so dependent on local testing practice, it is allowed to vary as a spatial logit-Gaussian process across county.  The hospital admission (hospitalization) rate for a particular county (or specific hospital) is also treated as a spatio-temporal random effect.  Hospital specific parameters determining transition rates from floor to ICU, discharge, etc. are also treated via a mixed effects model, with prior distributions informed by macro data and refined with hospital/region specific data.

This approach has been remarkably accurate and a linchpin in Mayo Clinic's response to managing the COVID-19 pandemic.  The model accurately predicted the case/hospitalization rise for Mayo Clinic's Arizona (MCA) and Florida (MCF) campuses weeks in advance of the actual case rise in the summer of 2020.  This gave them the notice necessary to prepare for the surges that took place.  Meanwhile it also accurately predicted that Mayo Clinic's Midwest hubs, Rochester (MCR), Southwest Minnesota (SWMN), Northwest Wisconsin (NWWI), and Southwest Wisconsin (SWWI) would experience a modest rise and a long plateau.  Hospital leadership understood that with this model in place they would have at least a couple of weeks lead time prior to a significant surge.  Thus, these sites were able to focus on how they could most effectively get back to caring for patients in a safe and effective manner without needing to worry as much about saving capacity for COVID-19 patients.  Once a fall surge was imminent in the Midwest, the model reacted quickly and again provided an accurate representation of how high hospital census could get.  This model has also been used by the state of Minnesota to help guide difficult policy decisions.  When many other models were predicting very large rises in hospital census in the state of MN as a whole during the fall surge, our model accurately predicted a significant, but much more modest rise and fall.  Remarkably it has remained very accurate through the entire course of the pandemic.  Finally, the model is also being used to consult industry partners on where cases might be high and how best to navigate their business in the COVID-19 climate.

The rest of the paper is laid out as follows. Section~\ref{sec:model} describes the technical formulation of the spatio-temporal model for regional infections, hospitalizations, and staff absences.
Section~\ref{sec:results} presents model some of the results of the model and discussion of various use cases to help guide decisions at Mayo Clinic and at the state of Minnesota.  Section~\ref{sec:conclusions} concludes.


\vspace{-.05in}
{\singlespacing
\section{A Bayesian Framework for County Level SIR Modeling}
}
\vspace{-.05in}
\label{sec:model}

\begin{figure}[t!]
\vspace{-.13in}
\centering
\caption{Model State Space.}
\vspace{-.15in}
\includegraphics[width=.61\textwidth]{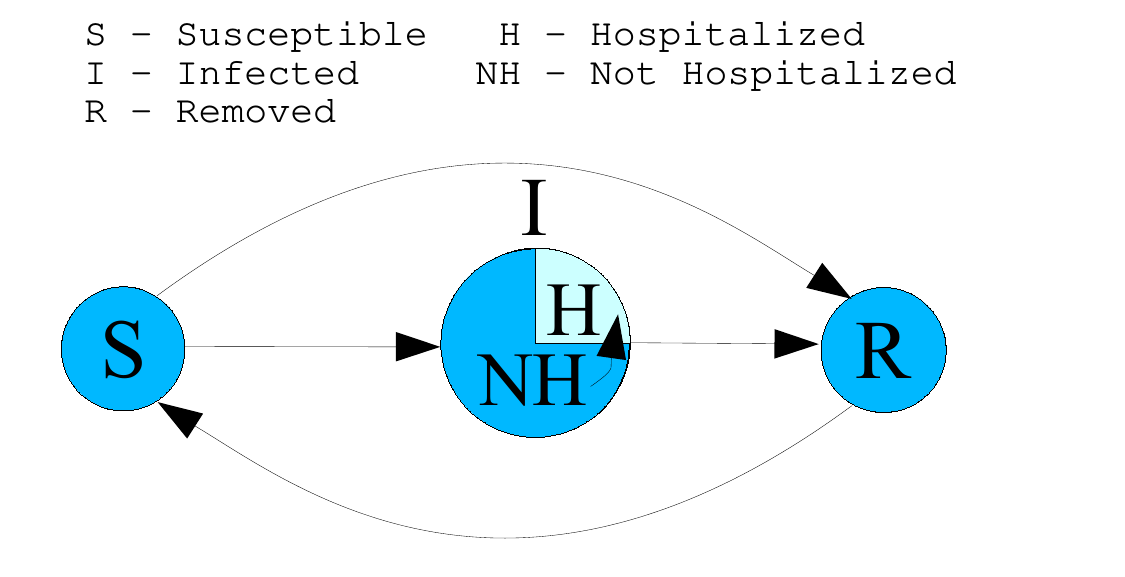}
\label{fig:state_space}
\vspace{-.1in}
\end{figure}

Figure~\ref{fig:state_space} provides a depiction of the state space for the compartmental disease model.  Individuals in each county start as susceptible, and then transition into infected or directly to removed (from vaccination).  There is also a transition from removed to susceptible again to allow for the impact of waning immunity and/or the possibility of breakthrough cases and new strains.  The infected pool can transition to hospitalization or to recovered (i.e., removed). Hospitalization is broken into general care (floor) or intensive care unit (ICU).  Transitions between these states are allowed as well as direct admission to the ICU.  Finally, hospitalized patients are allowed to transition into the removed state.

The model below allows for susceptible, infectious, and removed states at the US county level, whereas the additional states of hospital and ICU are treated on those becoming part of the infected class.  This essentially makes the assumption that hospitalization does not impact the rate of transition from infectious to removed.  This is likely not perfectly true, but it is a small percentage of those that would be affected and thus not have significant impact on overall future predictions.

\vspace{-.05in}
\subsection{Modeling Cases}
\vspace{-.05in}
\label{sec:case_model}

Let $\tY_{i,t}$ be the number of new cases reported in county $i$ on day $t$.  We also refer to this as the test$+$ count.  This is in contrast to $Y_{i,t}$ which is the actual number of new infections in county $i$ on day $t$, which is unknown but with the restriction $\tY_{i,t} \leq Y_{i,t}$.  The number of currently infected (or infectious) in county $i$ at time $t$ is $I_{i,t}$ and the number in the removed state are $R_{i,t}$.  Finally, $N_i$ is the number of people in residing in county $i$ and $S_{i,t}$ is the number of susceptible in county $i$ on day $t$.  In a traditional SIR model $S_{i,t} =  N_i - \sum_tY_{i,t}$, but that is not necessarily the case here since there is a path back to susceptible.

The model below first defines a model for how new infections $Y_{i,t}$ evolve from one day to the next in a discretized version of the true continuous time nature of the process.  In fact, the model below was motivated by discretized solution to a stochastic differential equations for a SIR model.
The model for number of test$+$ and total test volume then follows conditional on $Y_{i,t}$.

The model for number of new infections is 
\beq
Y_{i,t} \sim \mbox{Pois} \left( I_{i,t-1} B_{i,t-1} \frac{S_{i,t-1}}{N_i} \right),
\label{eq:Y}
\eeq
where
\beq
B_{i,t} = \exp\{\mu_{i,t} + \eps_{i,t} \},
\label{eq:B}
\eeq
The $\mu_{i,t}$ space-time Gaussian process over county and day and $\eps_{i,t} \stackrel{iid}{\sim} N(0, \sigma^2)$ are {\em iid} error to allow for noise in day counts.  The mean of the $\mu_{i,t}$ process is,
\beq
\E \mu_{i,t} = \phi_i + \bz_{i,t}' \balpha_i \bx_{i,t}'\bbeta,
\label{eq:mu}
\eeq
where $\bphi = [\phi_1, \dots, \phi_S]' \sim N(\mu_\phi, \bSigma_\phi)$ is a spatial random effect to allow the baseline level of spread rate to vary by county. The $\bz_{i,t}$ and $\bx_{i,t}$ are county level (possibly time varying) covariates vectors, to allow for day of the week effects, social distancing factors, or leading predictors such as mobility/proximity or Google trends, with corresponding spatial random effects $\balpha_i$ and fixed effects $\bbeta_j \sim N(0, \bS^2_\beta)$.  Denote each of the elements of $\balpha_i$ as $\balpha_i= [\alpha_{i,1}, \dots, \alpha_{i,K}]'$, and let $\balpha^{(k)}=[\alpha_{1,k}, \dots, \alpha_{S,k}]'$.  Each $\balpha^{(k)} \sim N(0,\bSigma_\alpha)$, independently over $k$. 

There are 3,109 counties in the contiguous US.  Since a traditional multivariate normal model requires ${\cal O}(N^3)$ operations for likelihood evaluation and/or realizations, we assume Gaussian Markov random Field (GMRF) model \citep{Rue05, storlie2017c} for the $\phi_i$ and $\balpha_i$ in order to allow for computational tractability.

The $\phi_i$ are assumed to be a GMRF with conditionally autoregressive (CAR) representation
\vspace{-.05in}\beq
\E(\phi_{i} \mid \phi_{-i}) = \mu_\phi + \rho_\phi \; \frac{\sum_{i' \in \cN_{i}} \left( \phi_{i} - \mu_\phi \right)}{N_{i}}
\label{eq:cond_mean}
\vspace{-.05in}\eeq
and precision
\vspace{-.05in}\bdm
\left[\Var(\phi_{i} \mid \phi_{-i}) \right]^{-1} = \frac{\tau_\phi} N_i,
\vspace{-.05in}\edm
where $\cN_{i}$ is the set of neighbors for county $i$ and $N_{i} = |\cN_{i}|$ is the number of neighbors to county $i$.

The CAR representation in (\ref{eq:cond_mean}) results in the following precision matrix $\bQ_\phi=\tau_\phi \bGamma(\rho_\phi)$ for $\bphi$,
\vspace{-.05in}\beq
(\bGamma(\rho))_{i',i} = \left\{
\begin{array}{ll}
  - \rho   &  \mbox{if $i' \in \cN_{i}$} \\
  N_i  &  \mbox{if $i' = i$} \\
  0 & \mbox{otherwise.}\\
\end{array}
\right. 
\label{eq:Qs}
\vspace{-.05in}\eeq

Each of the $\balpha^{(k)}$, $k=1,\dots,K$, are assumed to follow independent GMRF models entirely analogous to that for $\phi_i$ with precision $Q_{\alpha_k} = \tau_{\alpha_k} \bGamma(\rho_{\alpha_k})$ and mean $\mu_{\alpha_k}$.

The variance $\bSigma_\mu$ of the $\mu_{i,t}$ process is assumed to be that of a separable Gaussian Markov random Field (GMRF) model in space and time in order to allow for computational tractability.  Specifically, it is assumed that
\beq
\bSigma_\mu ^{-1} = \bQ_\mu = \bQ^{(s)}_{\mu} \otimes \bQ^{(t)}_{\mu}, 
\label{eq:Qmu}
\eeq
where $\bQ_{\mu,s} = \tau_\mu \bGamma(\rho^{(s)}_\mu)$ is the precision matrix over geographic space and taking the same form as that in (\ref{eq:Qs}), and $\bQ^{(t)}_{\mu} = \bOmega(\rho^{(t)}_\mu)$, where $\bOmega(\rho)$ is the precision matrix for autoregressive order 1, process with correlation parameter $\rho$ and unit variance to maintain identifiability of $\tau_\mu$.

The model for the number of new reported cases is governed by the number of new actual true cases and the amount of testing done per capita.  Specifically,
\bdm
\tY_{i,t} \sim \mbox{Binom}\left(Y_{i,t}, \pi_{i,t} \right),
\edm
where
\bdm
\mbox{logit}(\pi_{i,t}) = \tbeta_1 + \tbeta_2 \mbox{log}\left(\frac{M_{i,t}}{N_i} \times 10^5 \right)+ \teps_{i,t},
\edm
$M_{i,t}$ is the number of tests done on non-infected individuals, i.e., $\tY_{i,t} + M_{i,t}$ equals the total number of tests performed in county $i$ on day $t$, and $\teps_{i,t} \sim N(0,\tsigma^2)$ is {\em iid} error to allow for noise above and beyond the binomial distribution.

The number of infected individuals evolves through a much simpler model that assumes a geometric waiting time for individuals in the infected state, i.e.,
\beq
I_{i,t} = I_{i,t-1} - C_{i,t} + Y_{i,t},
\label{eq:I}
\eeq
where $C_{i,t}$ is the number of infected that transitioned into the removed class on day $t$ with
\bdm
C_{i,t} \sim \mbox{Binom}(I_{i-1,t}, \theta).
\edm
In the above, $1/\theta$ is the expected waiting time in the infected state and can be loosely interpreted as the expected length of SARS-CoV2 infection.  It is given the prior distribution, $\theta \sim \mbox{Beta}(10, 130)$, i.e., a mean of 13 days.

The number in the removed state $R_{i,t}$ follows an analogous model to $I_{i,t}$, assuming a geometric waiting time in the removed state (to transition back to susceptible), and a direct line from susceptible to removed via vaccination,
\beq
R_{i,t}= R_{i,t-1} - D_{i,t}+ C_{i,t}+ E_{i,t},
\label{eq:R}
\eeq
where (i) $D_{i,t}$ are the number in the removed class that transition back to susceptible, assumed to be geometric waiting time with prior distribution having mean of 1 year (2 years as a 95\% upper bound), and (ii) $E_{i,t}$ is the number of {\em effective} vaccinations that occurred in county $i$ on day $t-14$.  That is, vaccinations are assumed to not produce a transition into the removed state until two weeks after receiving it based on the biological science underlying the vaccines \citep{Polack2020,Baden2021}.  It is also assumed that not all vaccinations will transition into the removed class.  Specifically, the number of effective vaccinations $E_{i,t}$ is assumed to be proportional to the number of vaccinations received, $\tE_{i,t}$.  Specifically,
\bdm
D_{i,t} \sim \mbox{Binom}(R_{i,t-1}, \vartheta),
\edm
with the $\vartheta$ parameter given a Beta prior with $a=10$, $b=3,650$, i.e., a mean of 1/365 or expected immunity for 1 year.  This can and will be updated as more data and studies come out surrounding this issue.  The number of effective vaccinations is governed by,
\bdm
E_{i,t} \sim \mbox{Binom}(\tE_{i,t}, \pi).
\edm
The vaccination efficacy $\pi$ is given a prior distribution with mean 80\% allowing for the possibility that it is somewhere between 75 to 85\% based on results in several studies \citep{Tande2021,Baden2021,Polack2020,Voysey2021}.  This is slightly conservative as the clinical trials had reported higher efficacy at $>90$\%, but it is still a bit less clear how effective vaccines are at preventing asymptomatic infection and against newer variants.  Tande et al., 2021 report this later efficacy to be ~80\% in a recent study of a limited population in the state of Minnesota.  Thus, roughly 80\% of those that are vaccinated are assumed to move directly into the removed state after 2 weeks.  In this model, the two-week clock starts after the first dose regardless of vaccine type as even the two-dose Pfizer and Moderna vaccines have shown reasonable efficacy after even just a single dose \citep{Hunter2021,Chagla2021}.

Finally, the number in the susceptible state is simply determined by the constraint that $S_{i,t} = N_i - I_{i,t} - R_{i,t}$, imposed by the assumption that there is no migration across counties.  This assumption is clearly not true, but the population change in counties is small relative to population size during the course of a year or two.  Further, any increase in the spread rate of a neighboring county (possibly due to those living in one county and working in another, for example) is accounted for by the space time GMRF for $\mu_{i,t}$ in (\ref{eq:Qmu}).

The above formulation does not yet treat hospitalization.  These states are treated as special sub-states of those in the infected state. This is discussed in detail in Section~\ref{sec:hosp_model}.
The above formulation also does not specify an initial distribution.  Thus, for practical purposes the above process needs to be initialized with counts in each state at some initial time point $t_{i,0}$ for each county.  This was taken to be the day of the first reported case(s) in each county, and that on $t_{i,0}$ for a given county, $Y_{i,t_{i,0}} \def \tY_{i,t_{i,0}}$, $R_{i,t_{i,0}}=0$, and $I_{i,t_{i,0}}=\tY_{i,t_{i,0}}$.  The numbers on the first day of reported cases are typically small, and the process quickly ``forgets'' the effect of this initialization a few days later as it burns-in.

\vspace{-.05in}
\subsection{Time Varying Covariates and Prediction}
\vspace{-.05in}
\label{sec:time_covs}

It is typical to treat time varying predictors, e.g., mobility/proximity data, test volume, vaccinations, etc., as fixed when performing inference.  They could be treated this way in the formulation above, and the model calibration could be completed.  However, the prediction step requires that the value of such variables be known forward in time as well.  Thus, such variables must be treated as a random process that can also be predicted, and conditional on these predictions, predictions of new cases can then be made.  A similar strategy is discussed in \cite{Arik2020}.  Thus, any time varying predictors included below are assumed to come from a model analogous to that above in (\ref{eq:Qmu}), but with an additional product correlation across the dimension of the type of covariate, $k=1, \dots, K$.

We give special attention here to one particular time varying predictor, vaccinations, since it has a substantial impact on the future.  Predicting the number vaccinated in the future is done in the following manner.  Assume that the number of total vaccinations each day $\tE_{i,t}$ are Poisson with mean $\lambda_{i,t}$, where log$(\lambda_{i,t})$ is a GMRF in space and time with mean $\nu_i$. The mean $\nu_i$ is assumed to have been a linear ramp up since 01/01/2021 and has then been constant since 03/14/2021 for estimation purposes.  That is, future $\tE_{i,t}$ are assumed to look a lot like they did the past few weeks, with some potential for slow down or speed up.  The vaccinations are assumed to continue in this fashion until we reach a point of “saturation” where there is no longer demand for vaccine.  This is currently assumed to happen at some point between 50 and 75\% of the population vaccinated, a priori, with a mean of 60\%.  
Note that while the model does allow a transition back to susceptible once vaccinated, this transition is assumed to be relatively slow ($\sim$1 year), so this will have little impact on the four-month projections.  However, the role of novel variants to SARS-Cov-2 could alter this rate.  Further study is needed to estimate this transition probability in the coming months.  Making effective predictions beyond four months will also require a reasonable model for re-vaccination in late 2021 and subsequent years.

\vspace{-.15in}
\subsection{Modeling Hospitalization States}
\vspace{-.05in}
\label{sec:hosp_model}

The infected state has two sub-states: not-hospitalized (NH) and hospitalized (H).
Those who enter the the infected state are assumed to enter in the sub-state of NH, and are allowed to transition into the hospitalized state according to a geometric waiting time.  If they transition into the Removed state first, then they were not hospitalized during that stay in the infection state.  Let the number of new hospitalization admissions on day $t$ be $H_{i,t}$.  It is assumed that
\beq
H_{i,t} \sim \mbox{Pois}(I_{i,t} \omega_{i,t}),
\label{eq:omega}
\eeq
where $\omega_{i,t}$ follows an analogous model to that of $\mu_{i,t}$, namely a GMRF with precision $Q_\omega = \tau_\omega \bGamma(\rho^{(s)}_\omega) \bOmega(\rho^{(t)}_\omega)$.  The mean of the $\omega$ process also takes a similar form
\bdm
\E(\omega_{i,t}) = \phi_i^{(\omega)} + Z_{i,t}^{(\omega)} \balpha_{(\omega),i} + X_{i,t}^{(\omega)} \bbeta_{(\omega)}.
\edm

For a given hospital of interest, say {\em hospital s}, e.g., Mayo Rochester, we define a catchment area, (all counties with any significant inflow to that hospital) and a catchment rate parameter $alpha_{\omega,i,s}$ where it is assumed that any hospital admission of an individual from county $i$ has a $alpha_{\omega,i,s}$ probability of being admitted to hospital s.  Upon admission to hospital $s$ there is an $\ups_s$ probability of direct admit to ICU, and the transition from floor to ICU or discharge is governed by simple geometric waiting times.  Namely, let floor, ICU and discharge be states 1, 2, and 3 respectively.  Then $\psi_{s,l,m}$ defines the probability of a given patient in state $l$ moving from state $l$ to state $m$ on a given day.  The discharge state is absorbing, i.e., $psi_{s,3,3}=1$, but patients are allowed to go from ICU to floor and possibly back to ICU, etc., before discharge.

The $\ups_s$ and $\psi_{l,m}$ are given hierarchical logit-normal priors distributions to encourage but not require the $psi_{s,l,m}$ to be similar across hospitals in the set of hospitals of interest.  A hospital of interest in our case are our six geographically divided Mayo sites (Rochester, Florida, Arizona, South East MN, South West MN, North West WI, and South West WI).  For these sites, the number of admissions, direct ICU admission, transitions to and from ICU,  and discharges are known without error.  When making predictions at a macro level (e.g., predicting hospitalizations for the entire state of MN), we use the parent mean of the hierarchical distribution as an approximation to the ``within-hospital'' parameters.

\vspace{-.05in}
\subsection{Imputation of County Level Data}
\vspace{-.05in}
\label{sec:county_imp}

New cases are available at the county level through USA Facts \citep{usafacts}.  Total test volume and hospitalization data is only publicly available at the state level through COVID Tracking \citep{covidtracking} until 03/07/21, and also at CDC \citep{CDC} and HHS \citep{HHS} through the current day.  Thus, an imputation step is needed to produce county level hospital admissions $H_{i,t}$, and (COVID negative) test volume $M_{i,t}$, conditional on state level data (i.e., restricting that counties within a state sum to the state total).  This can be conveniently handled within the Bayesian framework by treating these as additional parameters and updating them conditional on other parameters and the state level data in a Gibbs step as in \cite{Storlie2019}.

\vspace{-.15in}
\subsection{MCMC Algorithm}
\vspace{-.05in}
\label{sec:MCMC}

Complete MCMC details are provided in the Supplementary Material.  However, an overview is provided here to illustrate the main idea behind some of the more complicated updates.  The MCMC routine is a typical hybrid Gibbs, Metropolis Hastings (MH) sampling scheme (e.g., see \cite{Kruschke2014}). The complete list of parameters in the model described in (\ref{eq:Y}) and (\ref{eq:omega}) are
\vspace{-.20in}\beq
\Theta = \left\{
\theta, \vartheta, \bmu, \left\{\balpha^{(k)}\right\}_{k=1}^K, \left\{\bbeta^{(j)}\right\}_{j=1}^J, \bY, \bomega, \left\{\balpha_{(\omega)}^{(k)}\right\}_{k=1}^K, \left\{\bbeta_{(\omega)}^{(j)}\right\}_{j=1}^J, \sigma^2, \bH, \bM, \right.\\[-.2in]
\nonumber
\eeq
\beq
\!\!\!\!\!\!\!\!\!\!\!\!\!\!\!\!\!\!\!\!\!\!\!\!\!\!\!\!\!\! \left.\tau_\phi, \rho_\phi, \tau_\mu,
\rho^{(s)}_\mu, \rho^{(t)}_\mu, \tau_M, \rho^{(s)}_M, \rho^{(t)}_M, \tau_H, \rho^{(s)}_H, \rho^{(t)}_H, \tbeta, \tsigma^2
\right\},
\label{eq:param_list}
\vspace{-.21in}\eeq

{\bf More details forthcoming...}

\vspace{-.2in}
\section{Analysis Results}
\vspace{-.05in}
\label{sec:results}

\vspace{-.15in}
\subsection{Posterior Summary}
\vspace{-.05in}

\begin{figure}[t!]
\vspace{-.0in}
\centering
\caption{Model calibration posterior distribution of several key model parameters.}
\vspace{-.15in}
\includegraphics[width=.32\textwidth, height=.20\textheight]{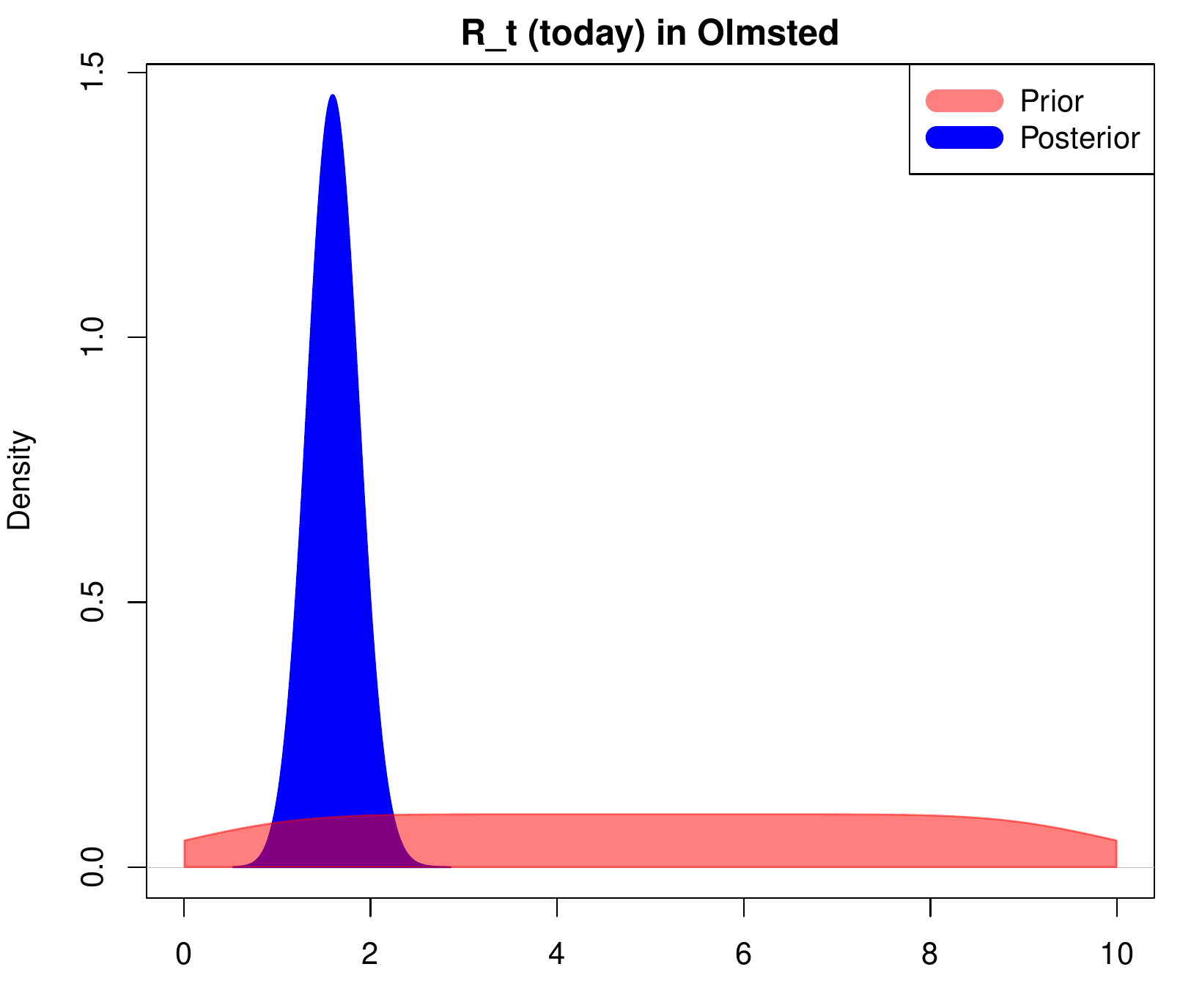}
\includegraphics[width=.32\textwidth, height=.20\textheight]{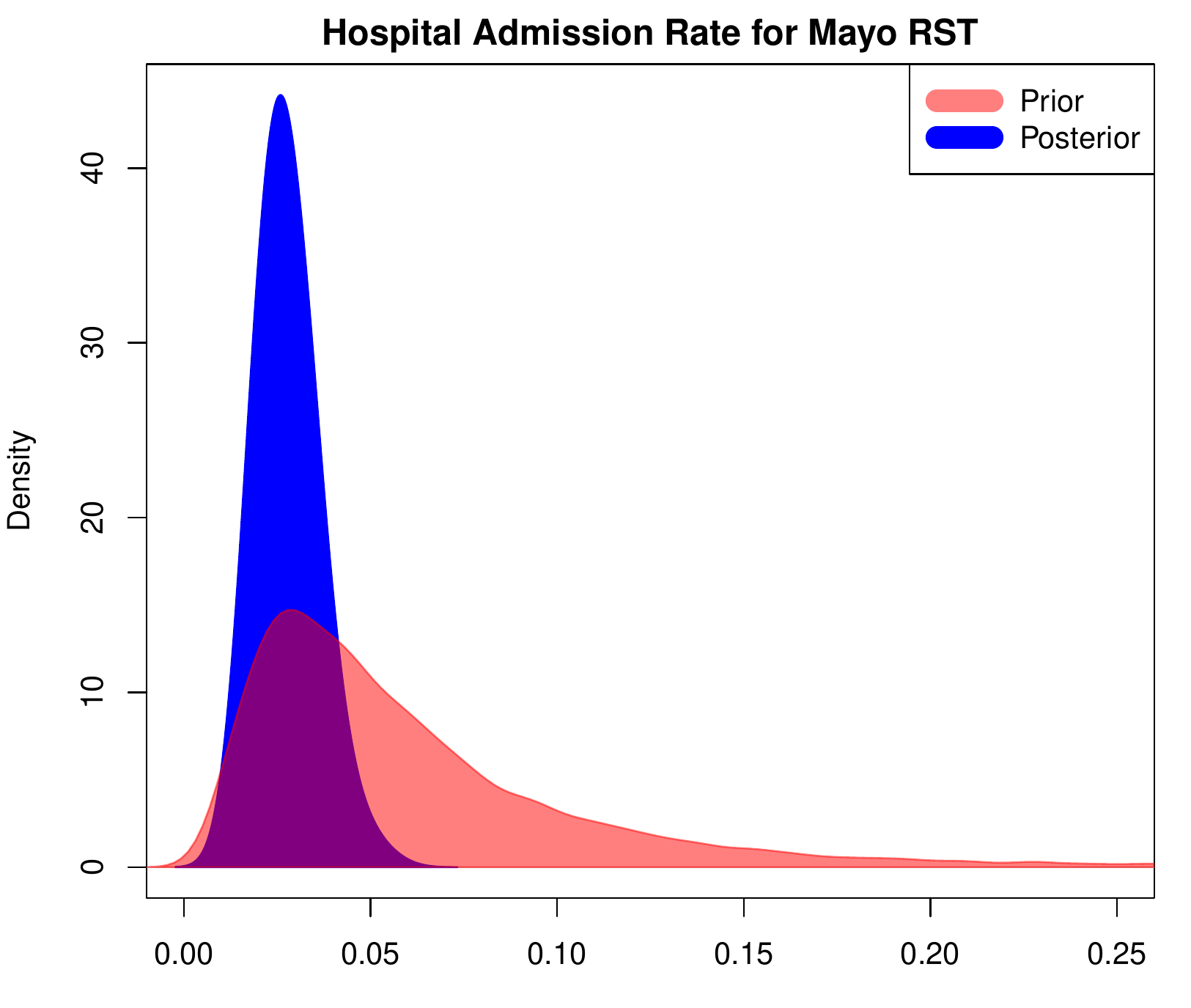}
\includegraphics[width=.32\textwidth, height=.20\textheight]{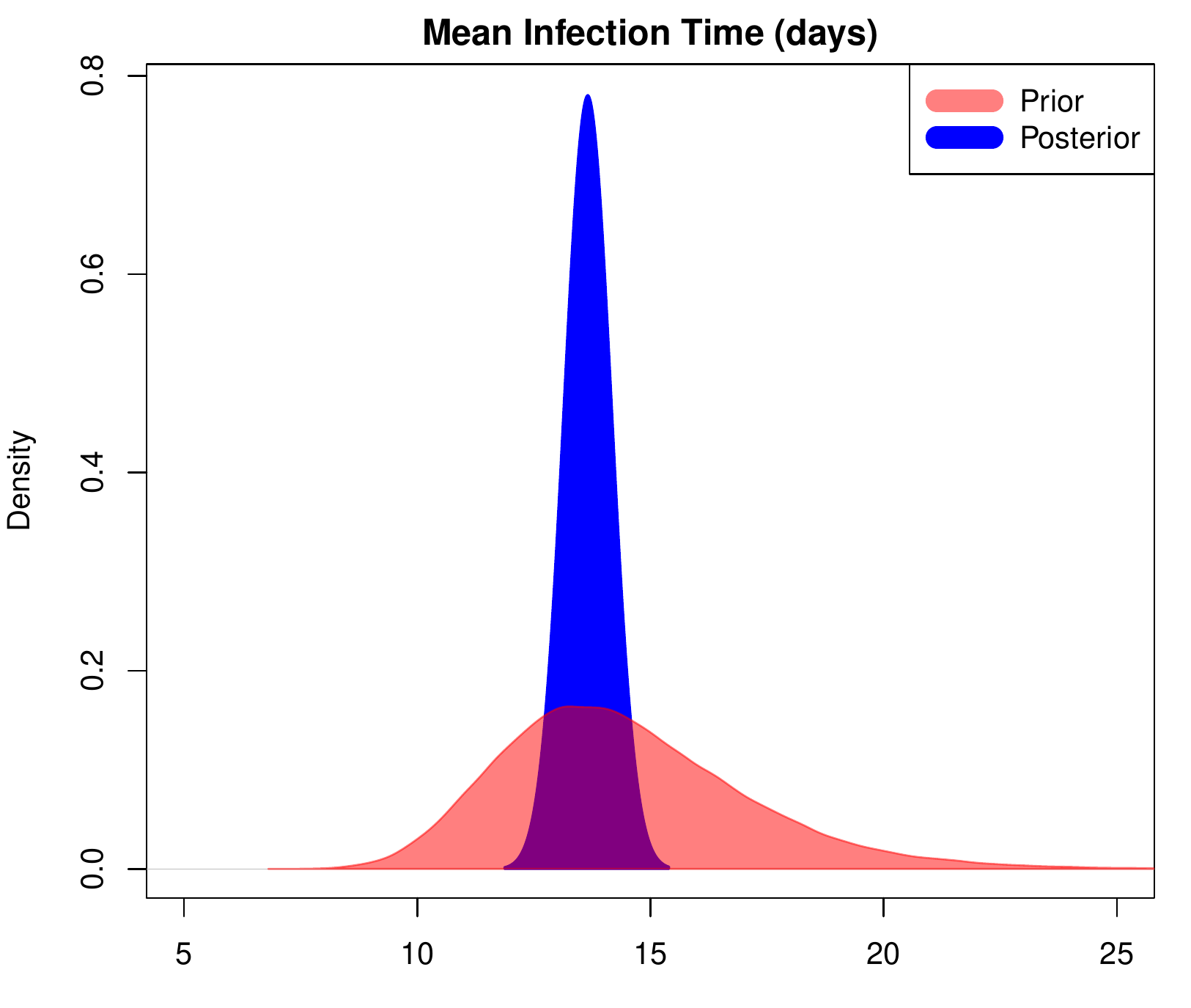}
\includegraphics[width=.32\textwidth, height=.20\textheight]{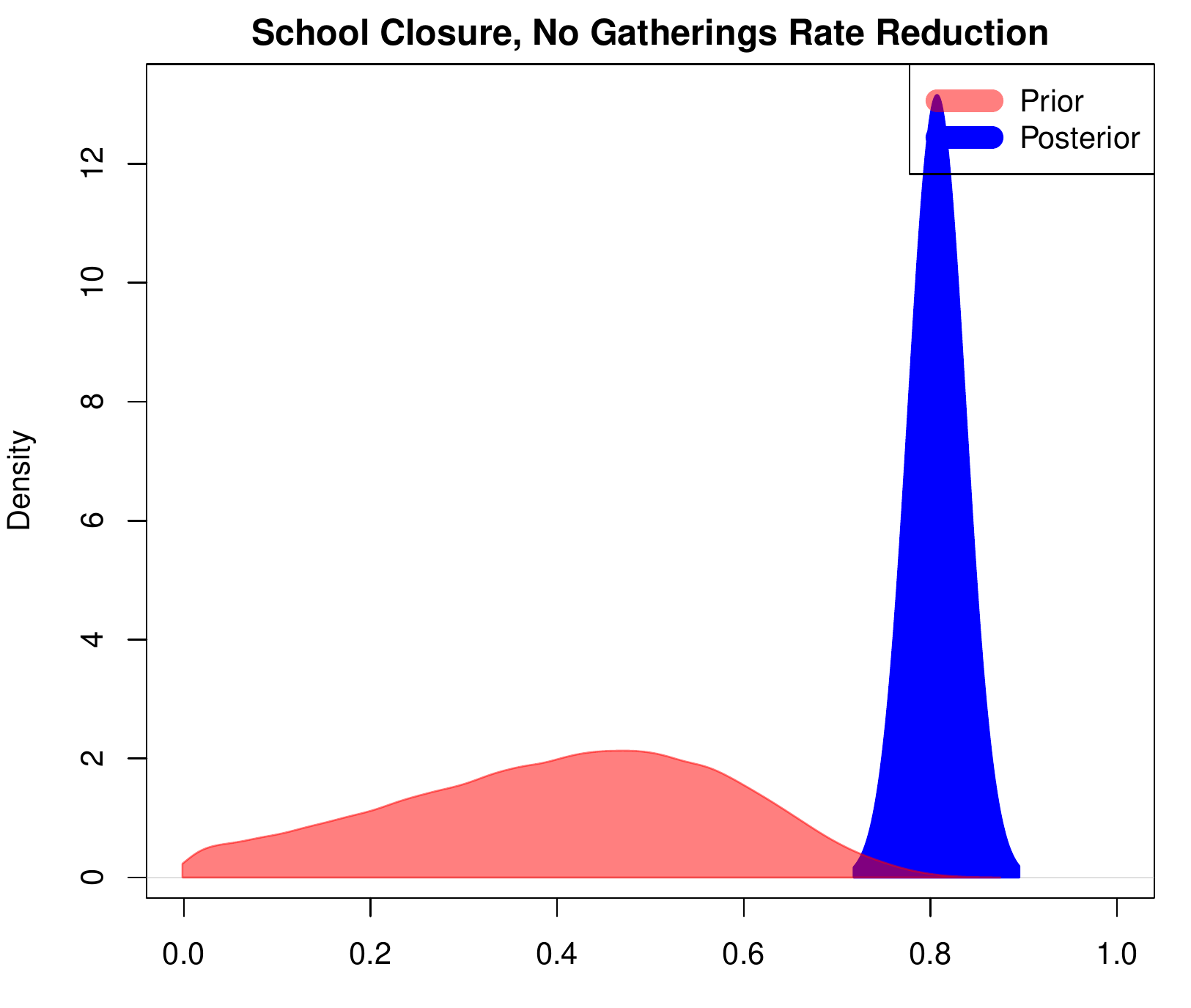}
\includegraphics[width=.32\textwidth, height=.20\textheight]{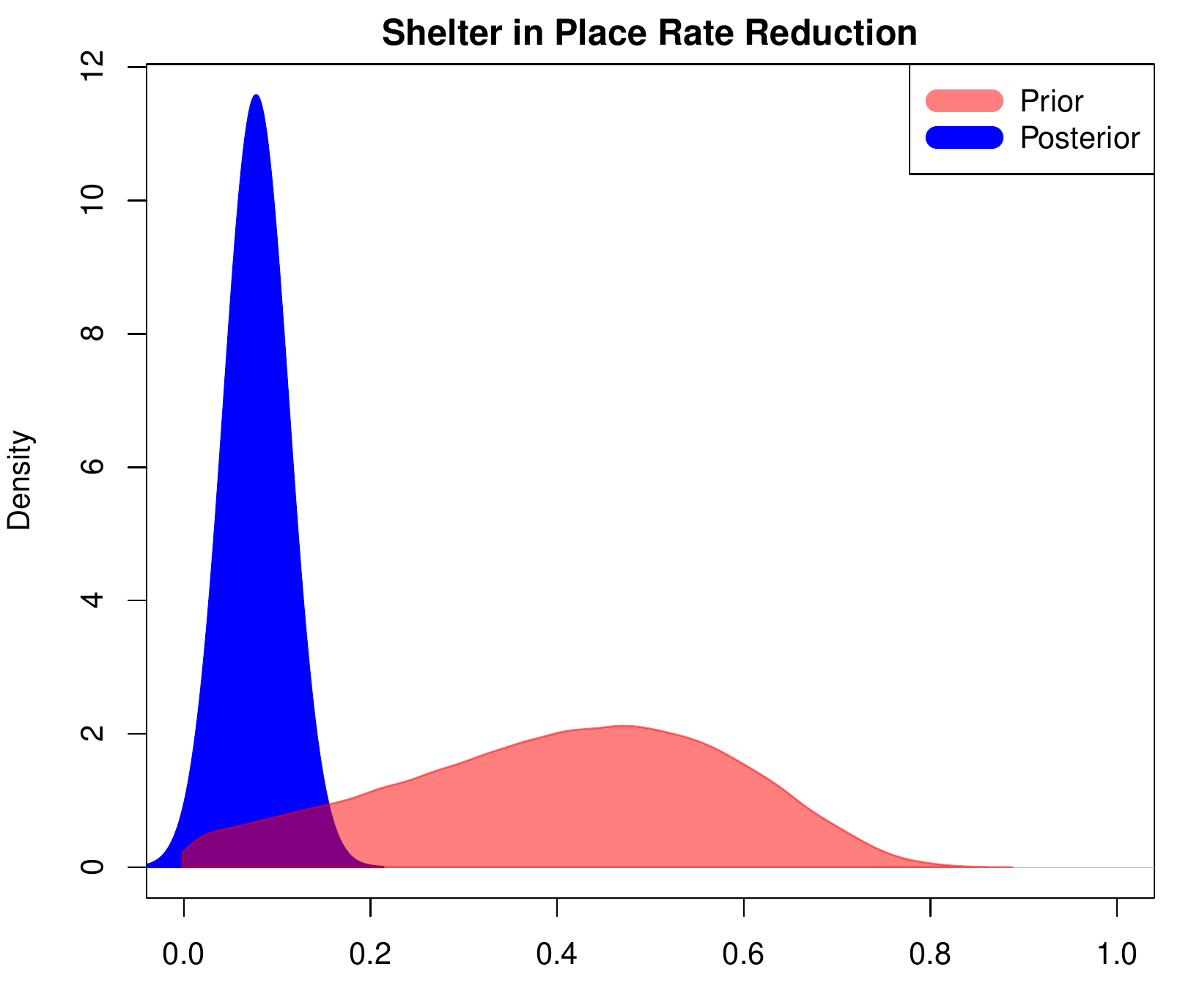}
\label{fig:param_posterior}
\vspace{-.0in}
\end{figure}

Table~\ref{tab:param_def} provides a high level description of a few key model parameters that are summarized in Figure~\ref{fig:param_posterior}.  Figure 4 provides an illustration of the Bayesian calibration process described in Figure~\ref{fig:bayes_UQ} put to work, refining the prior distributions to obtain a posterior representation of the uncertainty in the corresponding parameter values.

\begin{table}
\caption{Definition of several key model parameters}
\vspace{-.05in}
\begin{tabular}{p{1.9in}|p{3.4in}}
  {\bf Parameter} & {\bf Prior Distribution / Model Assumption}\\
  \hline
Rate of spread among infected in county $i$ at time $t$
& Space-time, log-Gaussian process over each US county and day\\
\hline
Proportion of people in county $i$ that are (test+) susceptible &
Spatial logit-Gaussian process across county\\
\hline
Rate reduction due to social distancing / social awareness &
log-Normal, mode 50\%, effective range (0,85\%) \\
\hline
Additional rate reduction due to stay-at-home order &
log-Normal, mode 50\%, effective range (0,85\%) \\
\hline
Hospitalization rate for each Mayo hospital &
Random effect process in time for each site with parent mean equal to respective state-level rate trend\\
\hline
\end{tabular}
\label{tab:param_def}
\end{table}

\vspace{-.15in}
\subsection{Mayo Clinic Hospitalization Prediction}
\vspace{-.05in}

These projections have been updated daily at six separate Mayo sites (Rochester, Florida, Arizona, South East MN, South West MN, North West WI, and South West WI) since April 2020.  Figure 5 displays the resulting forecast from propagating the posterior distributions through the model to predict hospital census for the next four weeks at Mayo Clinic Rochester (Floor and ICU) with a prediction date  of 02/03/2021 (four weeks prior to the time of this writing) along with the eventual observed data for the next four weeks superimposed.  The light grey curves are a sample of 10 posterior predictions to provide some context for the realm of possibilities that the model was expecting.  The interquartile range (IQR) is also provided (i.e., 50\% point-wise credible intervals).

\begin{figure}[t!]
\vspace{-.0in}
\centering
\caption{Four week future forecast for general care and ICU census at Mayo Clinic, Rochester, made on 02/03/2021 (the time of this writing), superimposed with the next four weeks of census data.}
\vspace{-.15in}
\includegraphics[width=.49\textwidth]{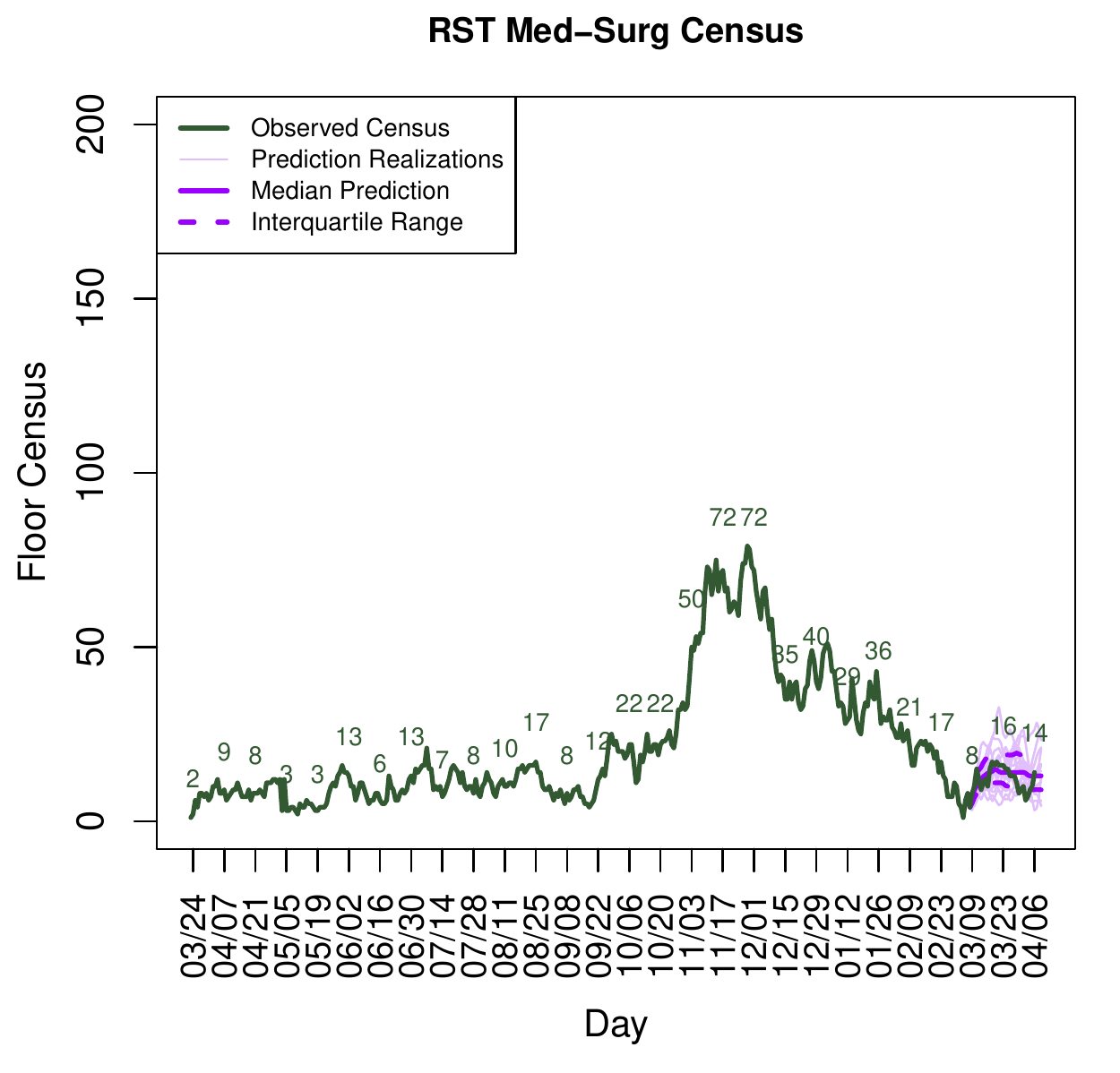}
\includegraphics[width=.49\textwidth]{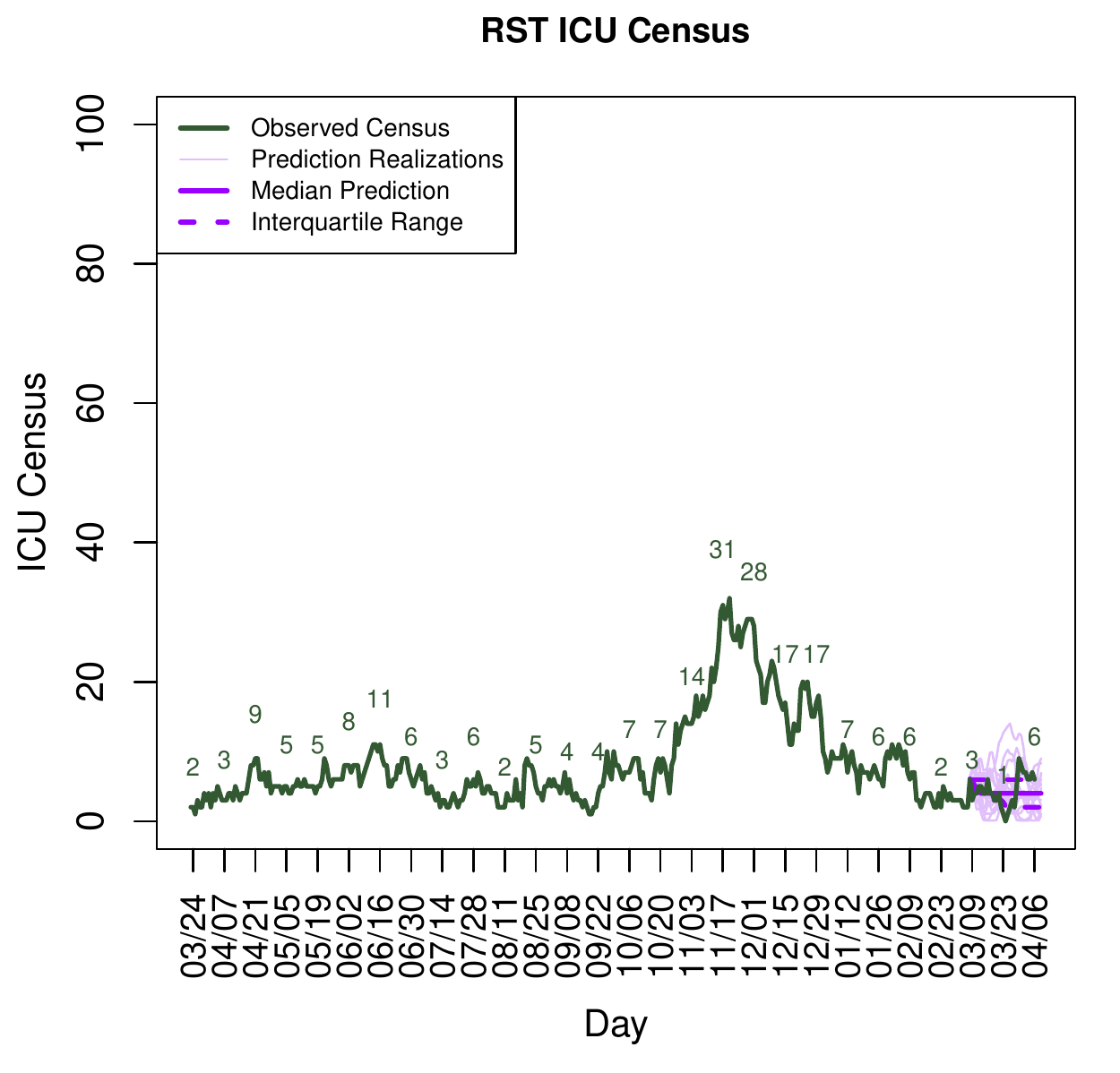}
\label{fig:RST_next4}
\vspace{-.0in}
\end{figure}

\begin{figure}[t!]
\vspace{-.0in}
\centering
\caption{Forecast for general care and ICU census for Mayo Clinic, Rochester, made on 11/01/2020, superimposed with subsequent observed data through 03/03/21 (the time of this writing).}
\vspace{-.15in}
\includegraphics[width=.49\textwidth]{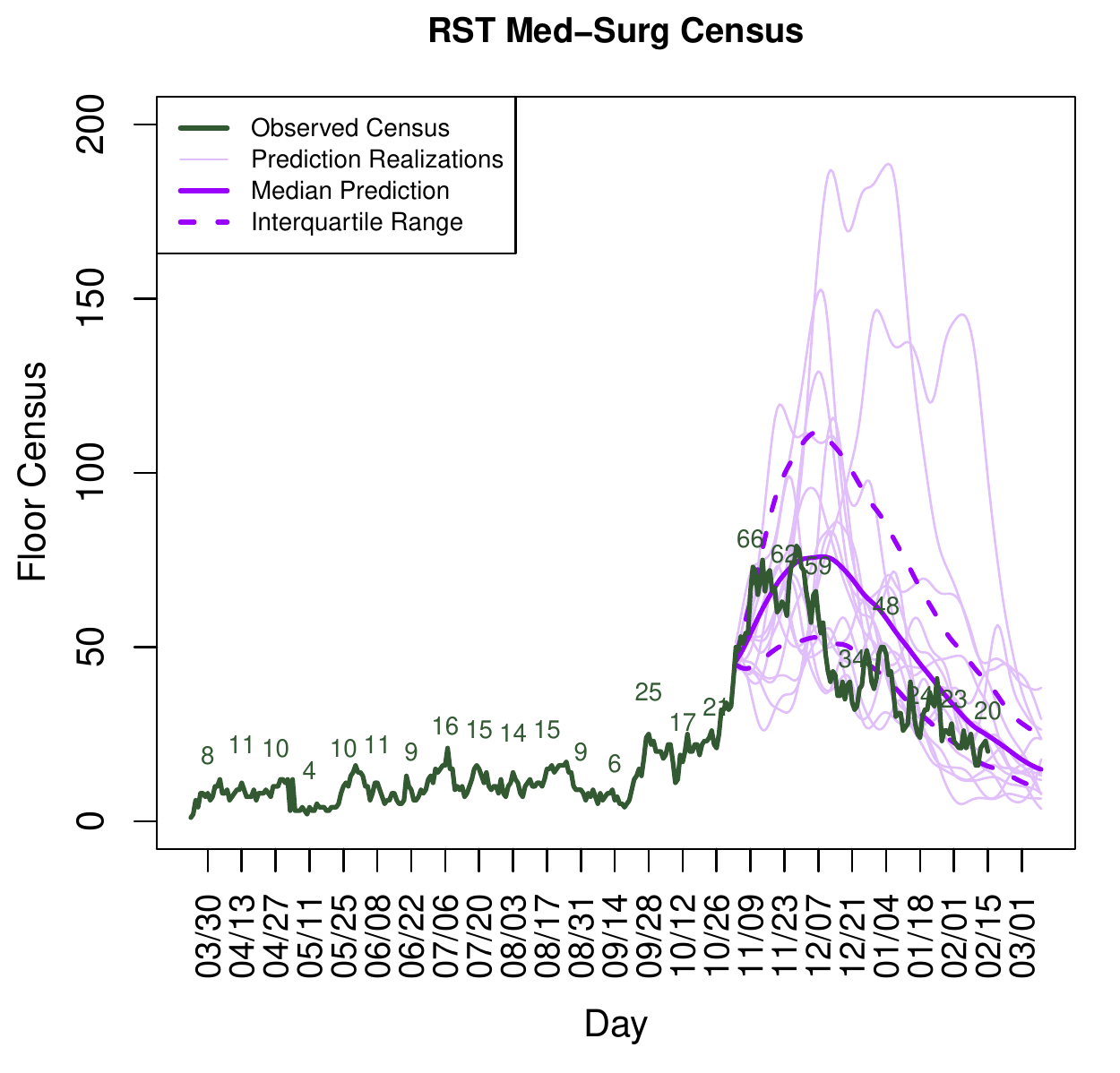}
\includegraphics[width=.49\textwidth]{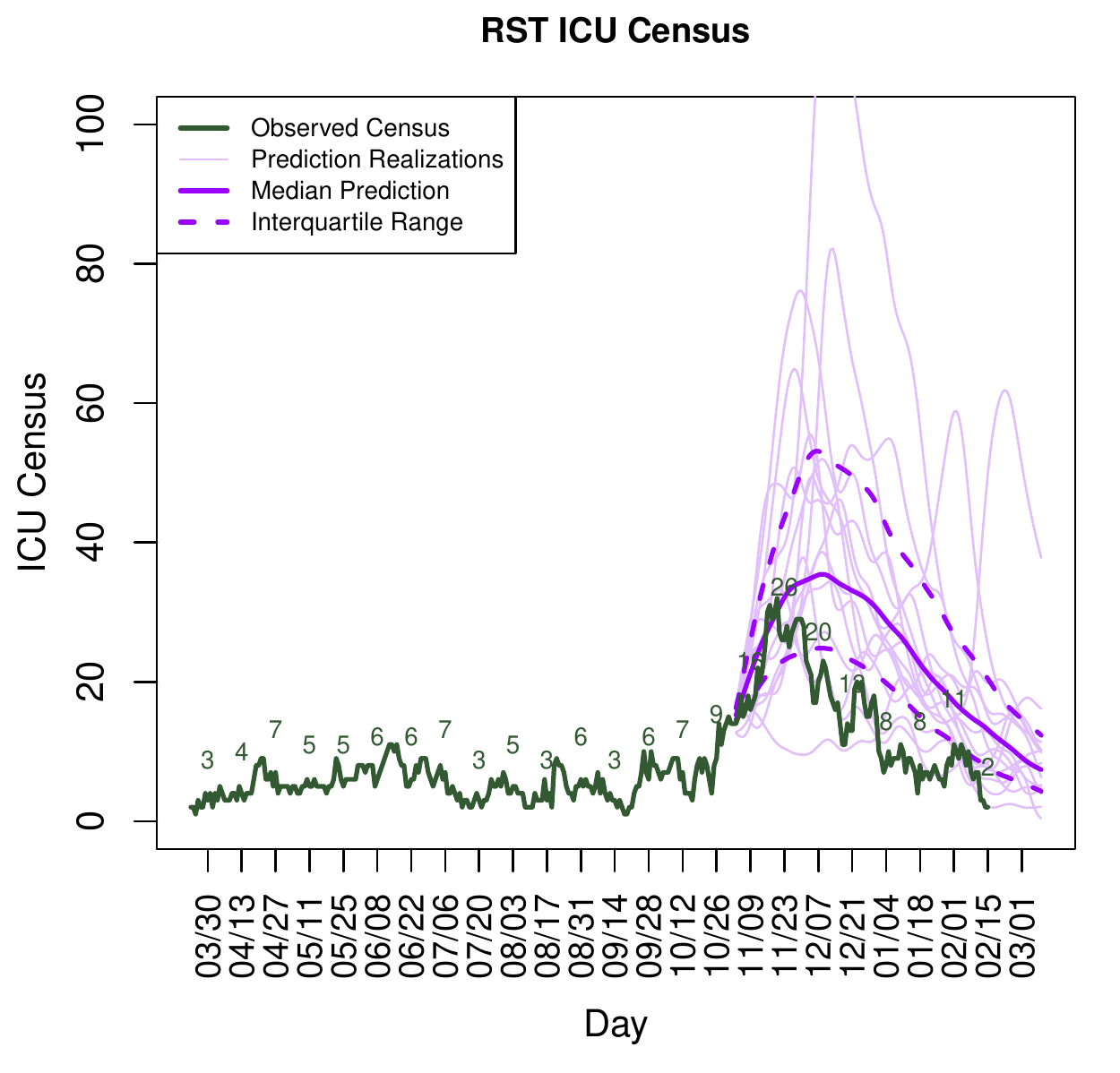}
\label{fig:RST_Nov_pred}
\vspace{-.0in}
\end{figure}

Of particular interest is how well the proposed model did to give warning during the COVID surge in the fall of 2020.  Back in early November 2020 the concern was at its peak in the US and the Midwest in particular, with cases rising and Thanksgiving/Christmas on the horizon.  Figure~\ref{fig:RST_Nov_pred} presents the prediction for Mayo Rochester hospital census (floor and ICU) made on 11/01/2020, along with the subsequent observed data through 03/03/2021 (the time of this writing).  it is clear that the approach presented here provided a median prediction and most model realizations very close to the ultimate reality.

\begin{figure}[t!]
\vspace{-.0in}
\centering
\caption{Four week future forecast for the probability that ICU census exceeds 50\% of capacity at each of the six Mayo Clinic sites (prediction from 11/01/2020)}
\vspace{-.15in}
\includegraphics[width=.9\textwidth]{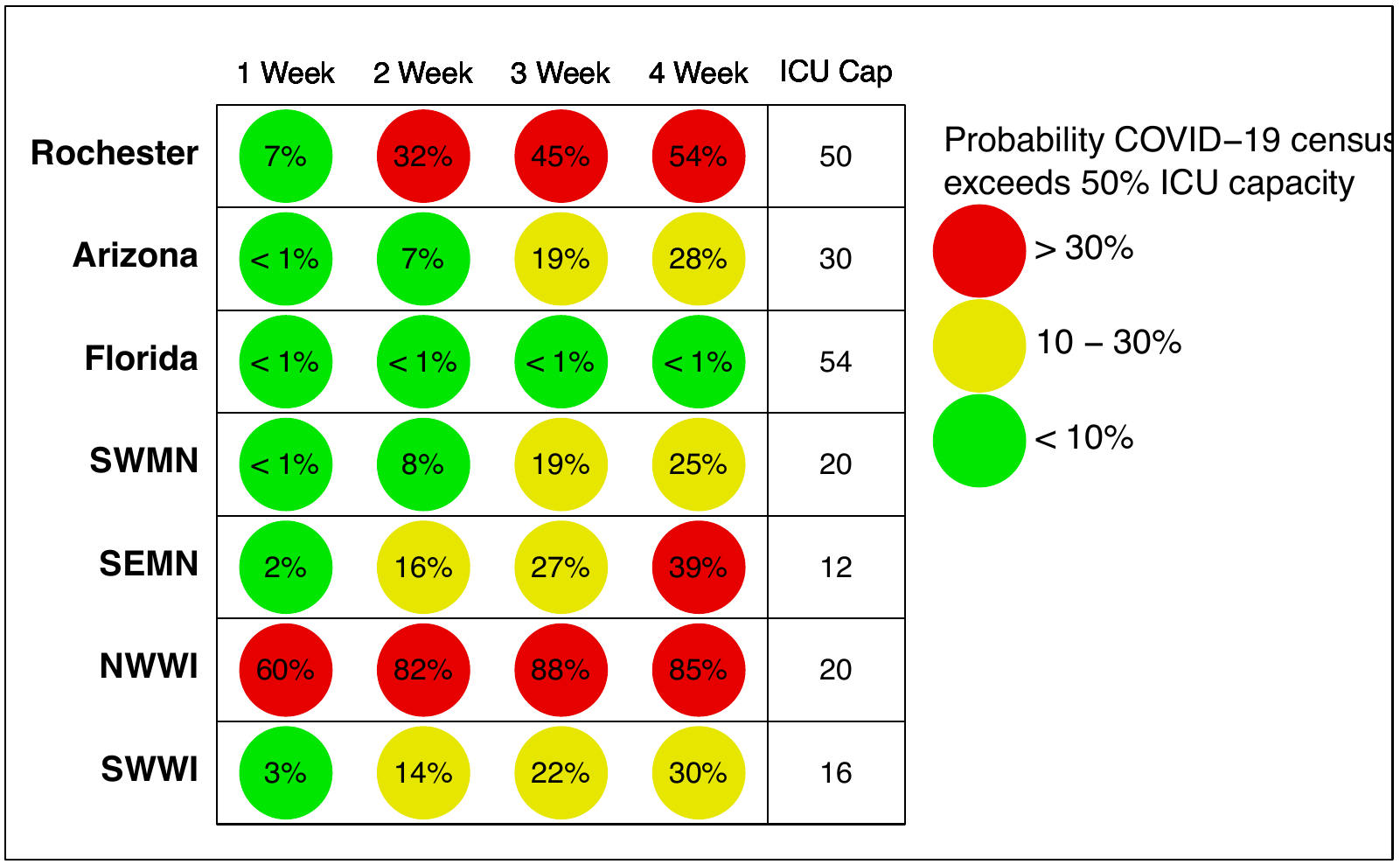}
\label{fig:stop_light}
\vspace{-.0in}
\end{figure}

\begin{figure}[t!]
\vspace{-.0in}
\centering
\caption{Four week future forecast for the number of staff absences (total and patient facing staff) for Mayo Clinic, Rochester (prediction made on 03/03/2021).}
\vspace{-.0in}
\includegraphics[width=.98\textwidth]{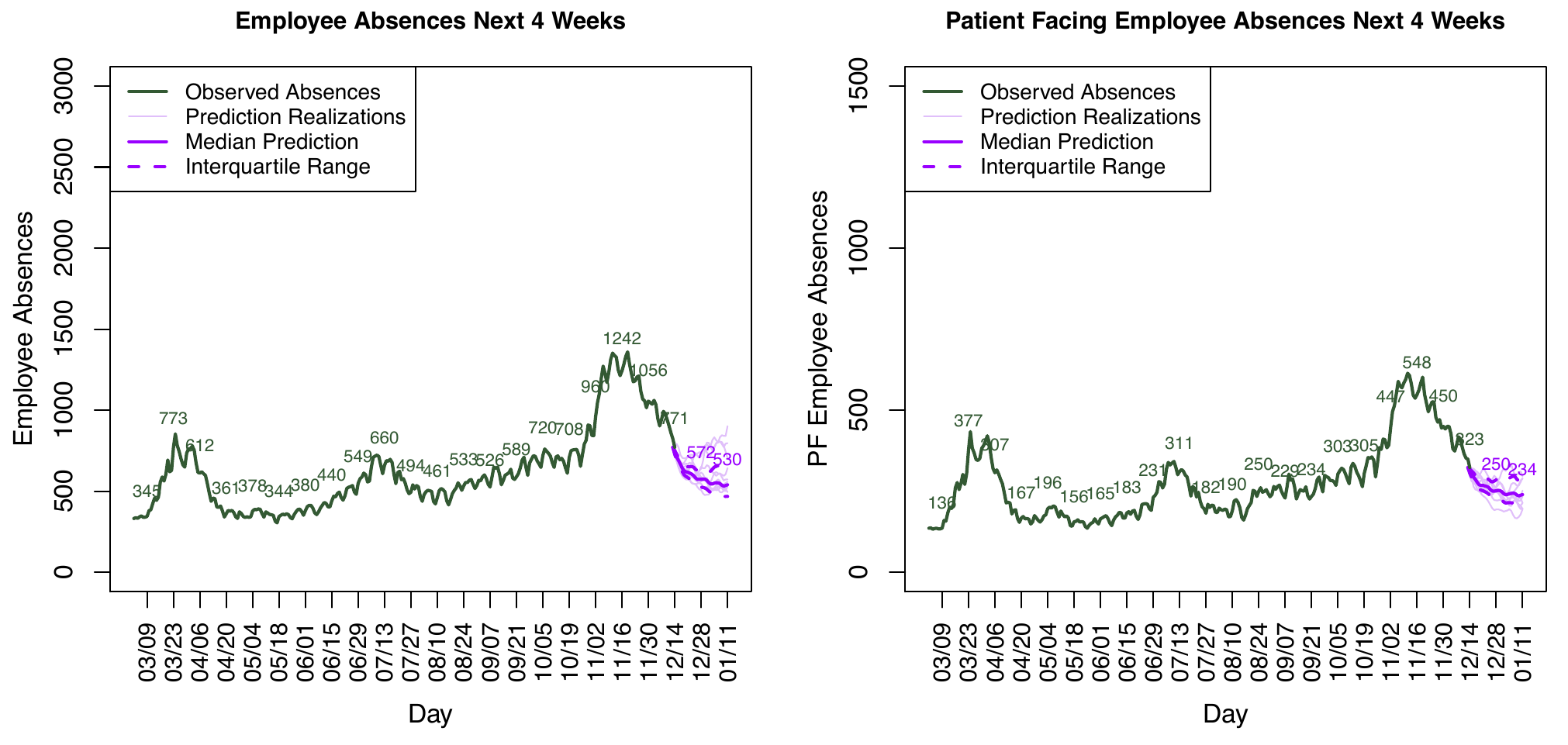}
\label{fig:rst_absence}
\vspace{-.0in}
\end{figure}

The Mayo leadership were particularly concerned about ICU capacity.  The results of the model can be summarized concisely into an ICU warning dashboard for leadership Figure~\ref{fig:stop_light} and predictions of Mayo staff absences made on 03/03/2020 are provided in Figure~\ref{fig:rst_absence}.

\vspace{-.15in}
\subsection{State of Minnesota Hospitalization Prediction}
\vspace{-.05in}

\begin{figure}[t!]
\vspace{-.0in}
\centering
\caption{Forecast for general care and ICU census in the entire state of Minnesota, made on 11/01/2020, superimposed with subsequent observed data through 03/03/21 (the time of this writing).}
\vspace{-.15in}
\includegraphics[width=.49\textwidth]{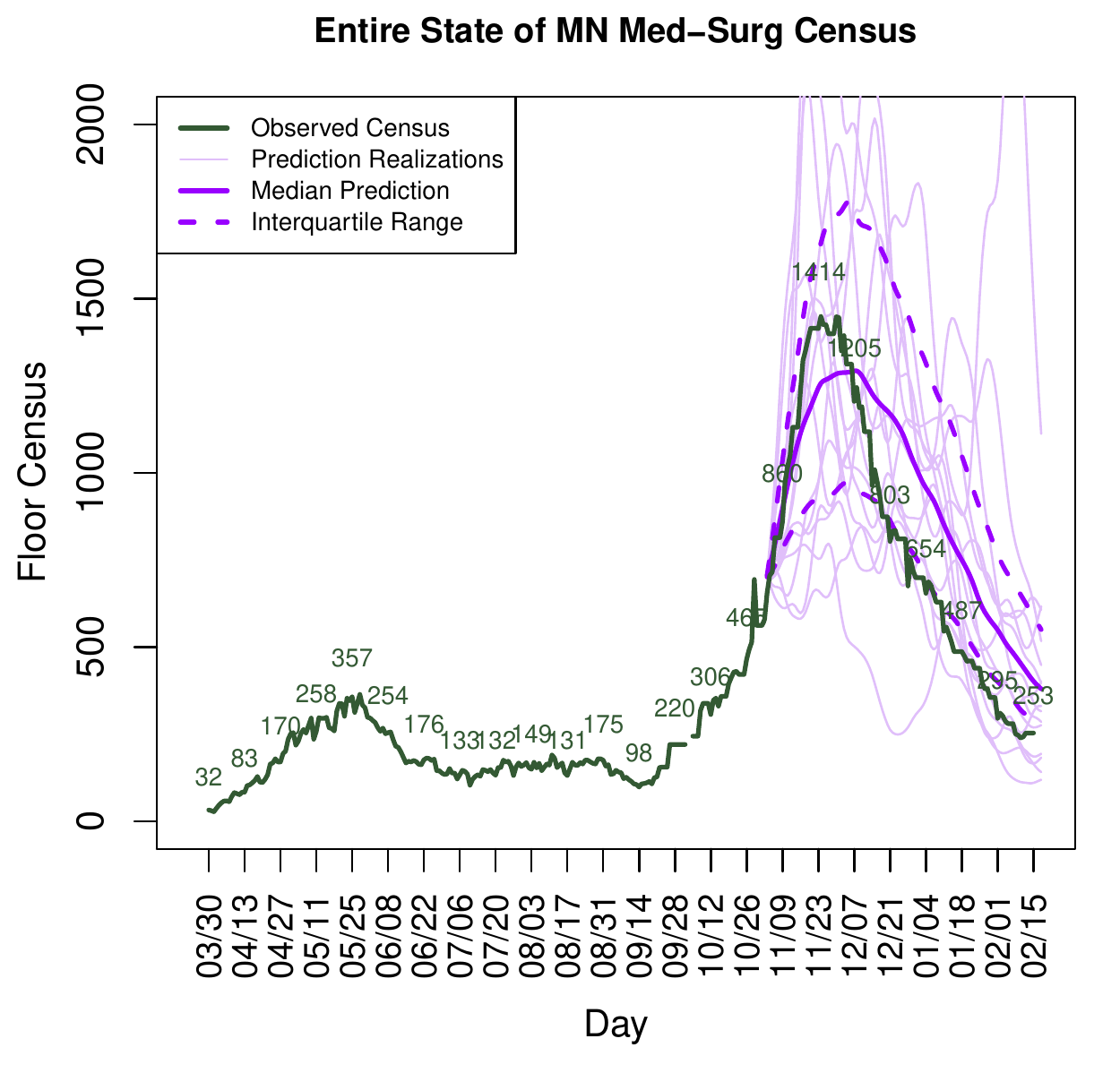}
\includegraphics[width=.49\textwidth]{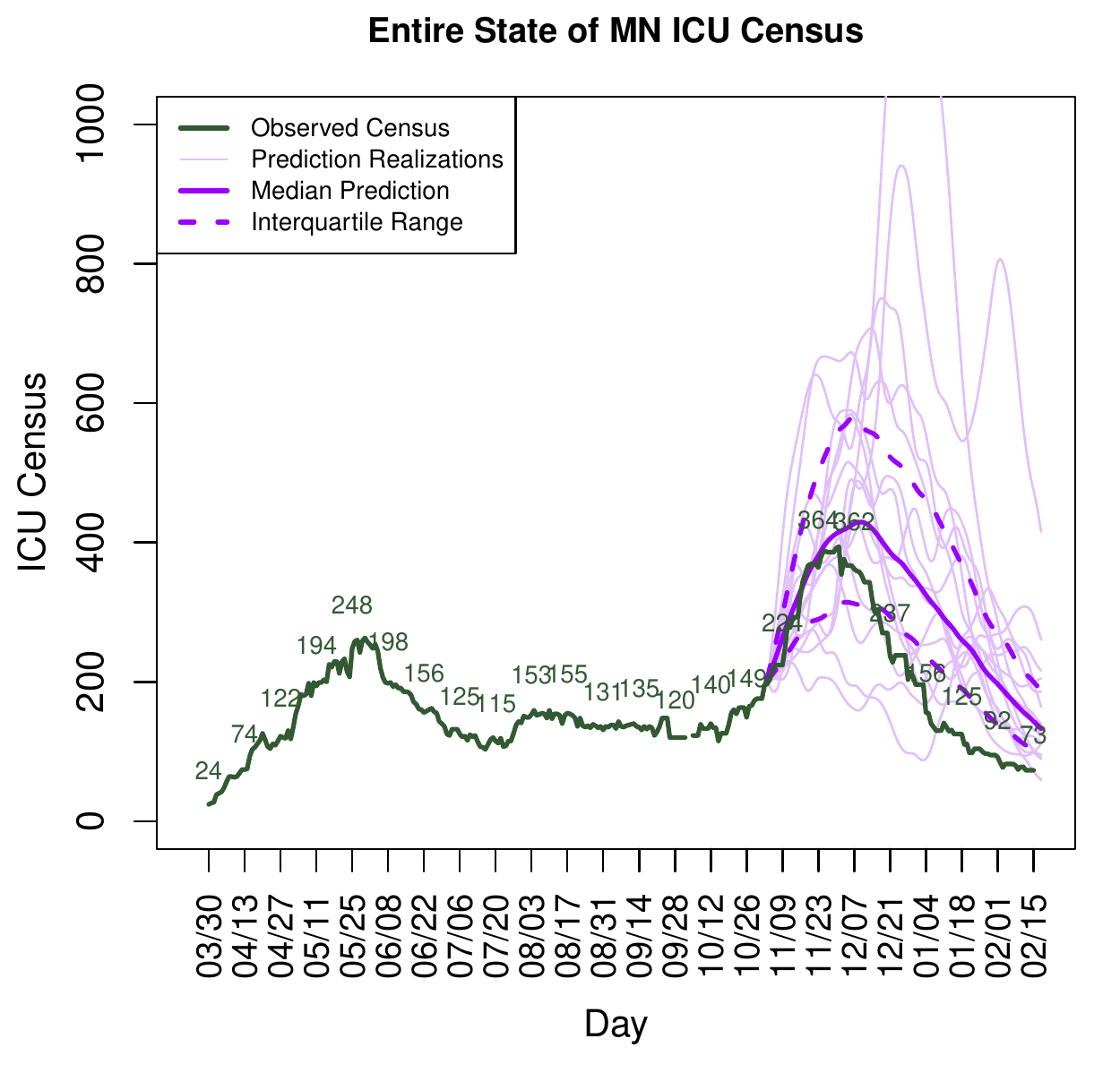}
\label{fig:MN_Nov_pred}
\vspace{-.1in}
\end{figure}

Finally, the approach can be used at a macro scale (e.g., state-level) to produce forecasts that can help policy makers.  Hospitalization forecasts were made for the entire state of Minnesota by treating the entire state as one big hospital.  Figure~\ref{fig:MN_Nov_pred}) provides the prediction made on 11/01/2020, again at the height of the concern about the beginning of the third wave of a COVID surge.  Once again, the proposed approach cam very close to predicting the actual rise and fall, while other modeling efforts were predicting much more pessimistic scenarios.

\vspace{-.15in}
\subsection{Prediction of Future Prevalence in US Counties}
\vspace{-.05in}

\begin{figure}[t!]
\vspace{-.0in}
\centering
\caption{Entire US, state-level view of the predictive model in use on the Mayo Clinic COVID webpage.}
\vspace{-.15in}
\includegraphics[width=.85\textwidth]{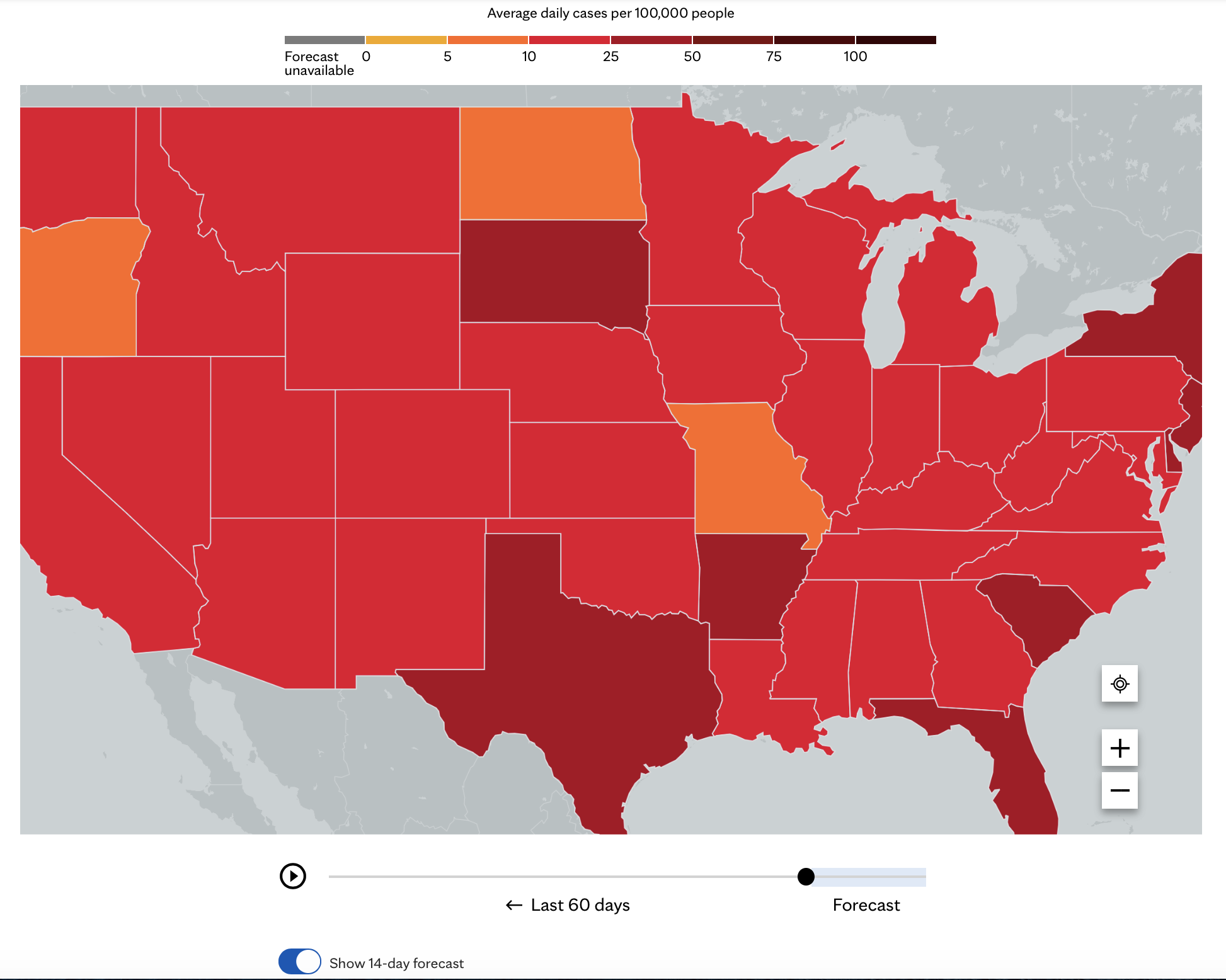}
\label{fig:state_map}
\vspace{-.1in}
\end{figure}

Finally, the model is now also being used in its rawest form to inform general COVID prevalence in different regions of the US.  For this purpose, we rely primarily on the 7-day case average as the metric for prevalence.  Figure~\ref{fig:state_map} displays a screen shot of the Mayo COVID website (\verb#https://www.mayoclinic.org/coronavirus-covid-19/map#, which now presents 14 day predictions of COVID-19 case activity at the county and state level.

\begin{figure}[t!]
\vspace{-.0in}
\centering
\caption{Forecast for daily cases and 7-day case average for the Minneapolis/ Saint Paul and surrounding area for the next eight weeks (prediction made on 04/06/21.}
\vspace{-.15in}
\includegraphics[width=.49\textwidth]{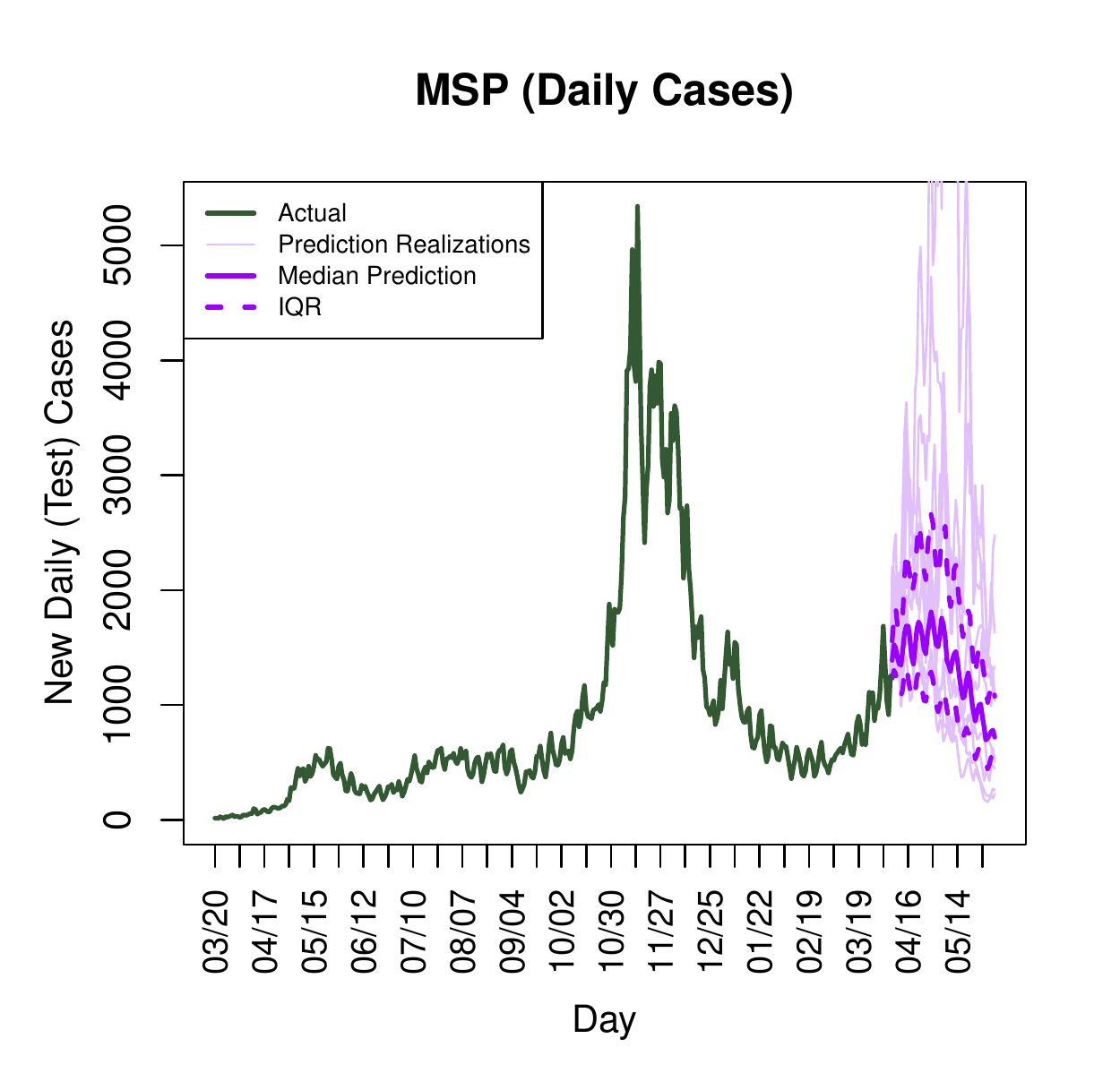}
\includegraphics[width=.49\textwidth]{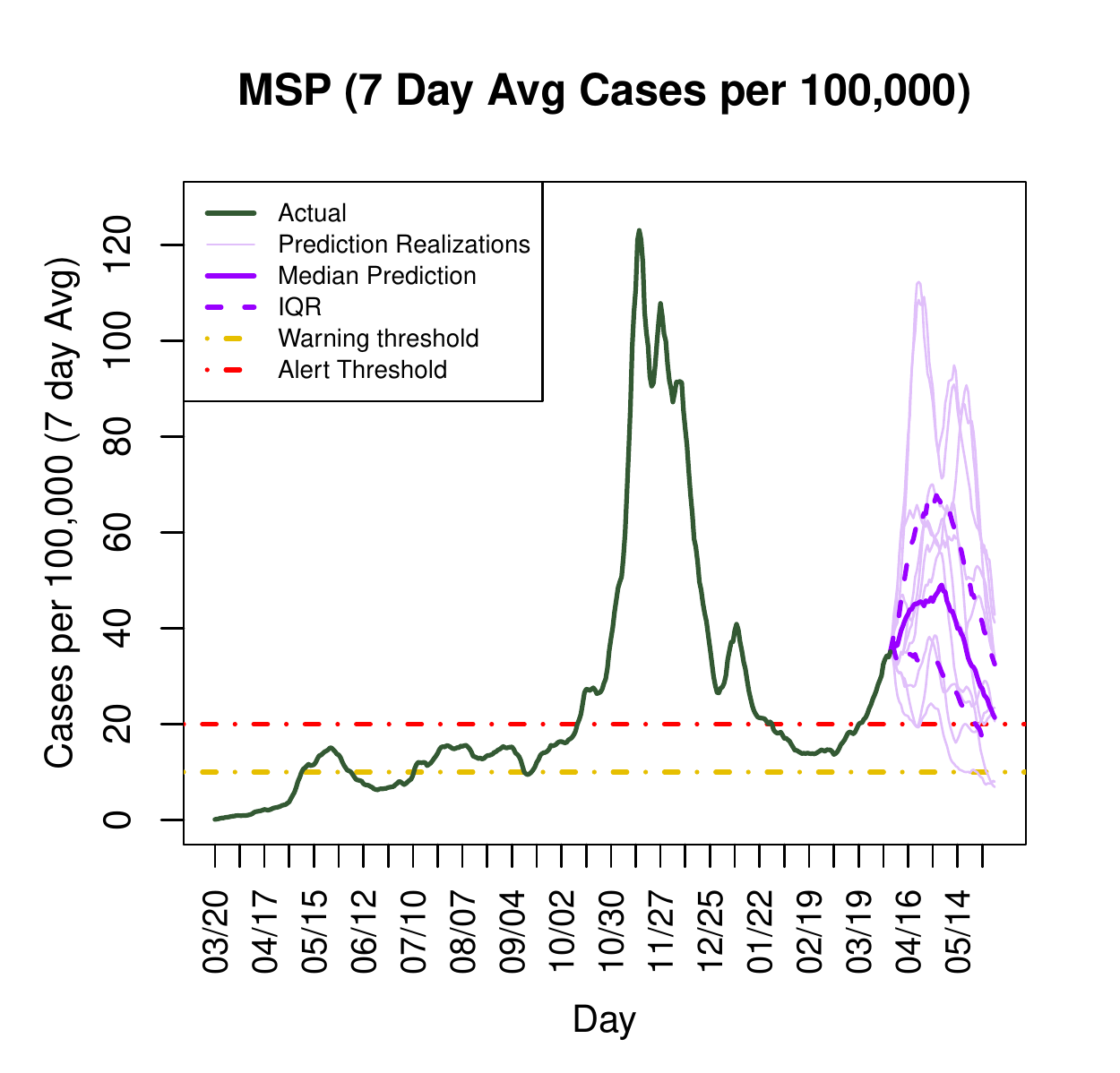}
\label{fig:MSP_prev}
\vspace{-.1in}
\end{figure}

The forecasts can also be rolled up to a metropolitan area to assist in planning for businesses or individuals.  Figure~\ref{fig:MSP_prev} displays this for the Minneapolis/ Saint Paul area, where it is expected to remain in a bit of a plateau around 15 cases per 100K in population.  There are some model realizations that do admit the possibility of another rise like that which was seen in the fall of 2020, however, a rise that substantial appears unlikely according to the model.

\vspace{-.25in}
\section{Conclusions \& Further Work}
\vspace{-.05in}
\label{sec:conclusions}

In this paper we have described a Bayesian compartmental model for COVID-19 infections and hospitalizations and how this model was used by Mayo Clinic and local government to help guide difficult decisions through the course of the pandemic.  The model served a very useful purpose to give fair warning of impending surges in cases hospitalizations and provided accurate forecasts of the magnitude of such surges.  It will continue to be used in this capacity in the coming months and years to help guide decisions in the face of further surges, which hopefully will never again be as pronounced as those seen in 2020.

While the approach has proven to be effective, still much work remains to continue to incorporate all of the changing characteristics of the pandemic.  Namely, mutations into new strains remain a significant wild card for future surges.  Also, more in-depth study into the length of immunity after infection (or vaccination) is needed to help inform prior distributions to limit the amount of confounding of parameters and make forecasts more precise.

Also, the model is currently driven primarily by case count and hospitalization trends.  These are lagging indicators of the community spread rate. However, it is clear that there is a strong relationship between internet search trends and social media data as leading indicators of new cases (cite Bydon).  Also, the demographics of those currently becoming infected can also play a pivotal role in predicting hospitalizations in the near term.  We intend to allow these data sources to enter into the above model as features that can affect the rate (of new cases and admissions) to improve the short term forecasting ability.  Google search trends are publicly available as an API in near real time.  The goal is to provide a health institution or government as much lead time as possible if there is an indication of a significant ramp up in community spread.

{\small
  \singlespacing
  \vspace{-.23in}
\bibliography{curt_ref}
\bibliographystyle{rss}
}

\end{document}